\documentclass[aps,prb,eqsecnum,twocolumn,showpacs,groupedaddress]{revtex4}
\usepackage{graphicx}
\usepackage{bbm}

\newcommand{\be}{\begin{eqnarray}}
\newcommand{\ee}{\end{eqnarray}}
\newcommand{\no}{\nonumber}
\newcommand{\Str}{{\rm Str}}
\newcommand{\tr}{{\rm Tr}}

\newcommand{\ba}{\begin{array}}
\newcommand{\ea}{\end{array}}
\newcommand{\bfr}{{\bf r}}
\newcommand{\bfrp}{{\bf r'}}

\newcommand{\bfp}{{\bf p}}
\newcommand{\bfq}{{\bf q}}

\begin{document}

\title{Supersymmetry for disordered systems with interaction}
\author{G. Schwiete$^{1}$ and K. B. Efetov$^{1,2}$}
\affiliation{$^{1}$ Theoretische Physik III,\\
Ruhr-Universit\"at Bochum, 44780 Bochum, Germany\\
$^{2}$L. D. Landau Institute for Theoretical Physics, 117940 Moscow, Russia\\
}
\date{\today }

\begin{abstract}
Considering disordered electron systems we suggest a scheme that allows us
to include an electron-electron interaction into a supermatrix $\sigma $%
-model. The method is based on replacing the initial model of interacting
electrons by a fully supersymmetric model. Although this replacement is not
exact, it is a good approximation for a weak short range interaction and
arbitrary disorder. The replacement makes the averaging over disorder and
further manipulations straightforward and we come to a supermatrix $\sigma $%
-model containing an interaction term. The structure of the model is rather
similar to the replica one, although the interaction term has a different
form. We study the model making perturbation theory and renormalization
group calculations. We check the renormalizability of the model in the first loop
approximation and in the first order in the interaction. In this limit we reproduce the renormalization group equations known from earlier works. We hope that the new supermatrix $\sigma $-model may become a
new tool for nonperturbative calculations for disordered systems with
interaction.
\end{abstract}

\maketitle



\section{\label{sec:introduction}INTRODUCTION}

Nonlinear $\sigma $-models serve as an important tool for the theoretical
description of disordered and chaotic systems. These models efficiently
describe physics at large distances and times, which is the most important
limit in the theory of localization, mesoscopic fluctuations, etc. The main
idea of the approach is to integrate out fast ``electronic degrees of
freedom'' and to reduce calculations to solving a model containing only slow
``diffusion modes''. This is the way how nonlinear $\sigma 
$-models are derived. The initial model of Ref.~\cite{Wegner79} was not completely satisfactory
because it contained formally divergent integrals. In order to avoid such
integrals one should either shift contours of integration in a complicated
way \cite{schaefer80} or start with a representation of the Green functions
in terms of functional integrals over anticommuting (Grassmann) variables 
\cite{Efetov80}.

Although these replica $\sigma $-models proved to be very convenient for
writing perturbation expansions in diffusion modes and renormalization group
equations, it became clear soon that they were not very useful in
nonperturbative calculations. As an attempt to improve the method, a
supersymmetric version of the field theory was formulated by Efetov\cite{Efetov82}. The supersymmetry method combines both the integration over
commuting and anticommuting variables and there is no need to use the
replica trick. It turned out that the supersymmetry method was very
efficient for studying various problems of disorder and chaos (for a review,
see, e.g., Ref.~\cite{Efetov97}).

The next step was to include electron-electron interactions and this was
done by Finkel'stein\cite{Finkelstein83} who generalized the $\sigma $-model of Ref. \cite{Efetov80} to interacting systems using the replica
approach and obtained a set of renormalization group equations. It is
interesting to mention that, for electron systems, the interaction can only be
included using the representation of Green functions in terms of
integrals over the anticommuting variables suggested in Ref. \cite{Efetov80}. The alternative representation of Refs. \cite{Wegner79,schaefer80} may
work for interacting bosons only. Needless to say that the $\sigma $-model
for interacting systems is considerably more complicated than that
suggested for noninteracting ones.

One more possibility to derive a $\sigma $-model for disordered systems is
based on the Keldysh technique \cite{Keldysh64}. As both the replica and
supersymmetry approaches, the Keldysh formalism enables one to average over the
disorder in the beginning of all calculations, which is the main step when
deriving the $\sigma $-models. The special form of the Keldysh time contour
automatically ensures that the weight denominator in the representation of
the Green functions via functional integrals is equal to unity, which is the
basis for performing the averaging over the disorder. The Keldysh formalism
for the disorder problems was first suggested in Ref.~\cite{Aronov85} and
later used in Ref. \cite{Horbach93} for derivation of a $\sigma$-model for
non-interacting systems. More recently, $\sigma $-models based on the
Keldysh formalism and the fermionic representation of Ref.\cite{Efetov80} were
derived for interacting systems \cite{Kamenev99,Chamon99}. These models, as
well as the previous ones, are applicable in the diffusive regime.

Although all the $\sigma $-models listed above are very similar as far as their structure and methods for perturbative expansions are concerned, they are quite different with respect to nonperturbative calculations. The simplest case where
this difference can be traced is the problem of the level statistics in a
disordered metal grain. In the $\sigma $-model language the level statistics
problem is equivalent to solving a zero-dimensional $\sigma $-model. For the
supersymmetric $\sigma $-model, one can reduce the calculation of the
level-level correlation function to the computation of a definite integral over $
2$ or $3$ variables (depending on the symmetries of the model). The latter
can be calculated without big difficulties and one comes to the famous
Wigner-Dyson statistics (see, e.g., Ref. \cite{Mehta91}). The method of calculations
based on supersymmetry can also be used for calculations for random
matrices \cite{Verbaarschot85a}.

The situation is quite different when trying to approach these problems with either 
replica or Keldysh $\sigma $-models. The first published observation of
difficulties encountered when calculating a proper integral over $N\times N$ matrices
($N$ is the replica number must be put to zero at the end) was
presented in Ref. \cite{Verbaarschot85}. Computation with the Keldysh $
\sigma $-model is not easier because one has time (or energy) as an
additional variable and integrals that appear for the zero-dimensional $
\sigma $-model are functional integrals rather than definite integrals over
several variables. Recently, several attempts have been undertaken in order
to improve this situation and to develop methods of nonperturbative
calculations for noninteracting replica\cite{Kamenev99a,Yurkevich99} and Keldysh \cite{Altland00} $\sigma $-models. Taking into account nontrivial saddle
points the authors of these works succeeded in reproducing oscillating
asymptotics of the Wigner-Dyson formulas. The latest achievement in this
direction is apparently the publication \cite{Kanzieper02} where the
level-level correlation function was explicitly derived for the unitary
ensemble using the replica approach.

In spite of the rather complicated calculations and limited success, the
attempts to perform nonperturbative calculations with the replica and
Keldysh $\sigma $-models were motivated by a probable extension to
interacting systems. Incorporation of interaction into the
supersymmetry scheme has not been considered and was generally believed to
be impossible. This opinion is, of course, not groundless. As we have
mentioned, in the presence of interaction, it becomes crucial whether one
considers fermions or bosons. The one particle picture that allows one to
consider the fermionic and bosonic representations on equal footing is
applicable for noninteracting particles only. In the supersymmetry method
one uses integrations over both fermionic and bosonic variables to describe,
e.g., a system of electrons and it is completely unclear how one can apply
this method for interacting particles.

There is no canonical way to incorporate interaction into a supersymmetric
model. In particular, it is hard to see how the normalization of the
partition function to unity could be achieved by adding bosonic variables
while keeping the full information encoded in the theory. In reality, we
cannot include an arbitrary interaction into the supersymmetric $\sigma $-model exactly. Therefore our goal is more modest and we restrict our
consideration to weak interactions.

Instead of trying to include interaction into the supersymmetry scheme we
propose an artificial model with interaction that is fully supersymmetric
and allows the standard treatment within the supersymmetry scheme (the
supersymmetry is violated only by source terms that allow us to calculate
different correlation functions). We justify the usefulness of this model by
comparing it with the initial electron model and find that, for a weak short
range interaction and an arbitrary disorder, the models are close to each
other. The comparison is performed by writing diagrams perturbative in the
interaction but exact in disorder. It is shown that on the level of a generalized
Hartree-Fock-approximation in the interaction (but exact in disorder)
the supersymmetric model we suggest and the initial electron model are
equivalent to each other.

Having demonstrated this equivalence we derive the $\sigma $-model using
integration over $4M$-component supervectors ($M$ is the number of Matsubara
frequencies, spin is included). Subsequently a supersymmetric $\sigma$-model is derived for
systems with unitary symmetry using the standard scheme of derivation
based on the averaging of the disorder, integration over supermatrices $Q$
and calculating the integrals in the saddle-point approximation, which fixes
the eigenvalues of $Q$. In the limit of vanishing interaction the resulting
model reduces to a form similar to the noninteracting $\sigma$-model, the main
difference being the size of the supermatrices that carry two additional
discrete frequency indices. For noninteracting systems the supersymmetry
technique proved to be a very powerful tool for the analysis of
nonperturbative effects related to disorder. The interesting question
arises how a weak interaction influences the physics in this regime. The
present model is intended to provide a starting point for tackling problems
of this kind.

Before addressing new problems, one should make sure that well-known
perturbative results of the theory of interacting electrons can be
reproduced with the supersymmetric $\sigma$-model derived. We address this
question in the present paper by studying the renormalization group
procedure introduced by Finkel'stein \cite{Finkelstein83} in the framework
of the replica $\sigma $-model. Of course, as our model is written for a
weak interaction, we cannot make a comparison for an arbitrary strength of
the interaction. Therefore, we restrict ourselves to writing renormalization
group equations in the first order in interaction. Under this restriction we
are able to follow the scheme of the renormalization of Ref. \cite{Finkelstein83} and demonstrate the renormalizability of the model. As a
result, we obtain renormalization group equations that agree with those
written in Ref. \cite{Finkelstein83} in the limit of small effective
interaction amplitudes.

This paper is organized as follows. Introducing the partition function of an
interacting electron gas with disorder in the Matsubara technique we first
consider the derivation of a $\sigma$-model for the theory without interaction
(Sec.~\ref{sec:functional}). Here we point out some technical
differences to the standard approach to this problem\cite{Efetov97}. Then we
turn to the interaction part in Sec.~\ref{sec:interactionpart}, motivating
the introduction of the supersymmetric model that is then studied in the
remainder of the paper. First we present a diagrammatic analysis (exact in
disorder), by which the model may be compared to the exact theory and the
range of applicability can be assessed. Then, a supersymmetric $\sigma$-model
is derived in Sec.~\ref{sec:smodel}. Contraction rules useful for
perturbation theory are given in Sec.~\ref{sec:perturbationtheory} and we consider 
the density-density correlation function as well as Altshuler-Aronov corrections\cite{Altshuler85} to conductivity. On these examples
we show how one can perform perturbative calculations for the
model derived. We demonstrate that in certain cases one can include also
higher order interaction effects. Next we derive the renormalization group
equations at lowest order in the interaction amplitudes (Sec.~\ref{sec:rengroup}) and also obtain the Altshuler-Aronov corrections \cite{Altshuler85} to the density of states.

\section{\label{sec:functional}Supersymmetric model without interaction}

The conventional supermatrix $\sigma $-model for noninteracting particles\cite{Efetov97} is written for two different frequencies. The interaction
makes all Matsubara frequencies important and we adjust the
supersymmetry formalism to this case. In order to understand the scheme
better we consider first the noninteracting case. Of course, the final
results of calculations with the effective model derived here must be the
same as those obtained in Ref. \cite{Efetov97} within the conventional
technique but there are some peculiarities in the intermediate steps.

The partition function of a disordered interacting fermion gas may be
written as a functional integral in the imaginary-time formalism \cite{Negele88} 
\be
Z=\int D\left( \chi ^{\ast },\chi \right) \;\mbox{e}^{-S}\label{eq:partition}. 
\ee
Here the integration is over Grassmann fields $\chi _{\sigma }(\mathbf{r},\tau )$ and $\chi _{\sigma }^{\ast }(\mathbf{r},\tau )$. The variables $\chi $ and $\chi ^{\ast }$ depend on the position $\mathbf{r}$, imaginary
time $\tau $ and the spin index $\sigma $ and obey the conditions 
\[
\chi _{\sigma }(\mathbf{r},\tau )=-\chi _{\sigma }(\mathbf{r},\tau +\beta
),\quad \chi _{\sigma }^{\ast }(\mathbf{r},\tau )=-\chi _{\sigma }^{\ast }(\mathbf{r},\tau +\beta ). 
\]
$\beta =T^{-1}$ is the inverse temperature. These constraints should be
respected when introducing the Fourier transform and we write
\[
\chi (\mathbf{r},n)=\sqrt{T}\int_{0}^{\beta }d\tau \,\mbox{e}^{i\omega
_{n}\tau }\chi (\mathbf{r},\tau ), 
\]
where $\omega _{n}=(2n+1)\pi T$ are fermionic Matsubara frequencies.

We decompose the action $S$ into three parts,
\[
S=S_{f}+S_{dis}+S_{int}. 
\]
The first term $S_{f}$ describes free electrons with a chemical potential $\mu $, 
\[
S_{f}=\int_{0}^{\beta }d\tau d\mathbf{r}\;\chi _{\sigma }^{\ast }(\mathbf{r}%
,\tau )\left( \partial _{\tau }+\frac{\hat{p}^{2}}{2m}-\mu \right) \chi
_{\sigma }(\mathbf{r},\tau ), 
\]%
where $m$ is the electron mass and summation over the spin index is implied.

The disorder part $S_{dis}$ describes the coupling to a static disorder
potential $U(\mathbf{r})$ 
\[
S_{dis}=\int_{0}^{\beta }d\tau d\mathbf{r}\,\chi _{\sigma }^{\ast }(\mathbf{r%
},\tau )U(\mathbf{r})\chi _{\sigma }(\mathbf{r},\tau ). 
\]%
When deriving the $\sigma $-model we will average over this potential
assuming that it is Gaussian and $\delta $-correlated 
\begin{equation}
\left\langle U\left( \mathbf{r}\right) \right\rangle =0\text{, \ \ \ }%
\left\langle U\left( \mathbf{r}\right) U\left( \mathbf{r}^{\prime }\right)
\right\rangle =\frac{1}{2\pi \tau \nu }\delta \left( \mathbf{r}-\mathbf{r}%
^{\prime }\right),  \label{e1}
\end{equation}%
where $\nu $ is the bare density of states at the Fermi surface and $\tau $
is the elastic scattering time.

By $S_0$ we denote the non-interacting part of the action 
\be
S_0=S_f+S_{dis}.
\ee
The interaction part $S_{int}$ is chosen in a conventional form 
\begin{eqnarray}
S_{int} &=&\frac{1}{2}\int_{0}^{\beta }d\tau d\mathbf{r}d\mathbf{r^{\prime }}%
\,\chi _{\alpha }^{\ast }(\mathbf{r},\tau )\chi _{\beta }^{\ast }(\mathbf{%
r^{\prime }},\tau )  \no \\
&&\qquad \times V_{0}(\mathbf{r}-\mathbf{r^{\prime }})\chi _{\beta }(\mathbf{%
r^{\prime }},\tau )\chi _{\alpha }(\mathbf{r},\tau ).\label{eq:2.6}
\end{eqnarray}%
where $V_{0}({\mathbf{r}}-\mathbf{r^{\prime }})$ is the bare potential of
the electron-electron interaction and the subscripts $\alpha $, $\beta $
stand for spin components.

In order to calculate correlation functions one should add to the action $S$
sources of a proper symmetry.

The aim of this section is to generalize the supersymmetry technique to all
Matsubara frequencies and express the partition function in terms of a
functional integral over supervectors. In order to arrive at a
supersymmetric form we should add bosonic variables. Some caution is
necessary to ensure convergence of the resulting Gaussian integrals, which
is very important for consideration of nonperturbative effects. To this end
we introduce new fermionic variables 
\begin{equation}
\chi _{n}^{\ast }\rightarrow i\chi _{n}^{\ast }\mathrm{sgn}\,(\omega _{n}),
\label{eq:2.12}
\end{equation}%
so that now $S_{0}$ takes the form 
\begin{eqnarray}
S_{0} &=&-i\sum_{n}\int d\mathbf{r}\;\chi _{\alpha }^{\ast }(\mathbf{r},n)\;%
\mathrm{sgn}\,(-\omega _{n})  \label{e0} \\
&&\qquad \times \left[ -i\omega _{n}+\frac{\hat{p}^{2}}{2m}-\mu +U(\mathbf{r}%
)\right] \chi _{\alpha }(\mathbf{r},n).\no
\end{eqnarray}%
To simplify the notation and to arrive at a familiar structure we arrange
the fermionic fields $\chi (\mathbf{r},n)$ with different Matsubara indices $%
n$ into vectors $\chi (\mathbf{r})$ in the following way 
\begin{equation}
\chi (\mathbf{r})=\left( 
\begin{array}{c}
\vdots \\ 
\chi (\mathbf{r},-2) \\ 
\chi (\mathbf{r},-1) \\ 
\chi (\mathbf{r},0) \\ 
\chi (\mathbf{r},1) \\ 
\chi (\mathbf{r},2) \\ 
\vdots%
\end{array}%
\right) .  \label{eq:2.13}
\end{equation}%
With the help of 
\begin{equation}
\Lambda _{nk}=\mathrm{sgn}\,(-\omega _{n})\delta _{nk}  \label{e12}
\end{equation}
the conjugate vector is defined as $\bar{\chi}(\mathbf{r})=\chi ^{\dagger }(%
\mathbf{r})\Lambda $, $\bar{\chi}(\mathbf{r},n)=\chi ^{\dagger }(\mathbf{r}%
,n)\mathrm{sgn}\,(-\omega _{n})$. We also use the frequency matrix $%
E_{nk}=-\omega _{n}\delta _{nk}$.

With these notations we rewrite the action $S_{0}$ in the form 
\begin{equation}
S_{0}=-i\int d\mathbf{r}\;\bar{\chi}_{\alpha }(\mathbf{r})\left[ iE+\hat{\xi}%
_{p}+U(\mathbf{r})\right] \chi _{\alpha }(\mathbf{r}),  \label{eq:2.15}
\end{equation}%
where $\hat{\xi}_{p}=\frac{\hat{p}^{2}}{2m}-\mu $.

Of course, when writing integrals over the anticommuting variables $\chi $ a
change of variables does not bring anything new. However, it is quite
important for introducing supervectors $\psi $ because one gets convergent
integrals over bosonic variables. We write these supervectors as follows 
\[
\psi (\mathbf{r})=\left( 
\begin{array}{c}
\vdots \\ 
\psi (\mathbf{r},-2) \\ 
\psi (\mathbf{r},-1) \\ 
\psi (\mathbf{r},0) \\ 
\psi (\mathbf{r},1) \\ 
\vdots%
\end{array}%
\right) ,\qquad \psi (\mathbf{r},n)=\left( 
\begin{array}{c}
\chi (\mathbf{r},n) \\ 
S(\mathbf{r},n)%
\end{array}%
\right) , 
\]%
where $S\left( \mathbf{r,}n\right) ,$ $S^{\ast }\left( \mathbf{r},n\right) $
are bosonic variables. Now we replace 
\[
\chi \rightarrow \psi ,\qquad \bar{\chi}\rightarrow \bar{\psi}=\psi
^{\dagger }\Lambda 
\]%
in $S_{0}$ to find the supersymmetric action $S_{0}^{\prime }$ 
\begin{equation}
S_{0}^{\prime }=-i\int d\mathbf{r}\;\bar{\psi}_{\alpha }(\mathbf{r})\left[
iE+\hat{\xi}_{p}+U(\mathbf{r})\right] \psi _{\alpha }(\mathbf{r}).
\label{e4}
\end{equation}%
With this preparation a $\sigma$-model can be constructed for the
noninteracting theory following standard considerations as described, for
example, in Ref.\cite{Efetov97}. For simplicity, we consider the ensemble
with the unitary symmetry, assuming that a small magnetic field suppresses
Cooperon modes. Then, it is not necessary to further double the size of the
supervectors. As a consequence, after the disorder averaging the resulting
quartic term may be decoupled with the help of a $4M\times 4M$ ($M$ is the
number of Matsubara frequencies) supermatrix fields $Q(\mathbf{r})$ with two
spin and two Matsubara indices 
\begin{eqnarray}
&&\mbox{e}^{-\frac{1}{4\pi \nu \tau }\int d\mathbf{r}\left( \bar{\psi}(%
\mathbf{r})\psi (\mathbf{r})\right) ^{2}}  \label{e10} \\
&=&\int DQ\;\mbox{e}^{-\frac{1}{2\tau }\int d\mathbf{r}\;\bar{\psi}(\mathbf{r%
})Q(\mathbf{r})\psi (\mathbf{r})-\frac{\pi \nu }{4\tau }\int d\mathbf{r}\;%
\mathrm{Str}Q^{2}(\mathbf{r})}.\qquad  \nonumber
\end{eqnarray}%
For the definition of the supertrace operation \ ``\textrm{Str'' }we refer
to Ref.~\cite{Efetov97}. In contrast to the orthogonal case no special
caution is necessary for the decoupling, since Cooperon modes are frozen.
The saddle point equation reads 
\begin{eqnarray}
&&Q(\mathbf{r})=\frac{1}{\pi \nu }\;\left\langle \psi (\mathbf{r})\bar{\psi}(%
\mathbf{r})\right\rangle =\frac{1}{\pi \nu }g(\mathbf{r},\mathbf{r}), \\
&&\left( iE+\hat{\xi}_{p}+\frac{1}{2\tau }Q(\mathbf{r})\right) \;g(\mathbf{r}%
,\mathbf{r^{\prime }})=i\delta (\mathbf{r}-\mathbf{r^{\prime }}),\label{eq:sp}
\end{eqnarray}%
and is solved by $Q=\Lambda $. In complete analogy to the standard case\cite%
{Efetov97} one derives the $\sigma$-model 
\begin{eqnarray}
Z &=&\int DQ\;\mbox{e}^{-F} \\
F &=&\frac{\pi \nu }{4}\;\int d\mathbf{r}\;\mathrm{Str}\big[D(\nabla
Q)^{2}-4EQ\big].
\end{eqnarray}%
The supermatrix field $Q(\mathbf{r})$ obeys the nonlinearity constraint $%
Q^{2}(\mathbf{r})=1$ and further $Q^{\dagger }(\mathbf{r})=KQ(\mathbf{r})K$. 
$D=v_{F}^{2}\tau /d$, where $d$ is the dimension, is the classical diffusion
coefficient. Here and in the following we adopt the convention that all
internal indices that are not displayed explicitly are summed over as part
of the supertrace operation. Repeated indices are also summed over. The
validity of the model is restricted to the diffusive regime, momenta are
restricted by the condition $ql\ll 1$ and temperature by $T\tau \ll 1$.

\section{Interaction part}

\label{sec:interactionpart} In the previous derivation of the $\sigma $%
-model for noninteracting particles containing all Matsubara frequencies we
closely followed a corresponding part of the paper \cite{Finkelstein83}. Now
we are to include the interaction. Unfortunately, unlike the fermionic replica
approach of Ref. \cite{Finkelstein83}, the supersymmetry does not allow to
include the interaction in a simple way and one cannot further follow the
scheme of Ref. \cite{Finkelstein83}.

Instead of trying to derive a $\sigma $-model from the original electron
model with interaction we follow another root, namely, we replace the
original electron model with interaction by an artificial supersymmetric
one. There are two requirements for this new model: it should not be very
different from the original one and it should be treatable within the
supersymmetry scheme. We have succeeded in constructing such a model that
gives the same correlation functions as the original one in the limit of a
weak interaction. Due to its supersymmetric form the derivation of the $%
\sigma $-model can be performed using the standard scheme.

We start our discussion writing the original interaction term $S_{int}$, Eq.
(\ref{eq:2.6}), in a slightly different form using the transformation, Eq. (%
\ref{eq:2.12}) and the vectors $\chi ,\bar{\chi}$, Eq. (\ref{eq:2.13}), 
\begin{eqnarray}
S_{int} &=&-\frac{T}{2}\int d\mathbf{r}d\mathbf{r^{\prime }}\;\left( \bar{%
\chi}^{\alpha }(\mathbf{r})\Delta ^{j}\chi ^{\alpha }(\mathbf{r})\right)
\label{Sint1} \\
&&\qquad \times V_{0}(\mathbf{r}-\mathbf{r}^{\prime })\left( \bar{\chi}%
^{\beta }(\mathbf{r}^{\prime })\Delta ^{-j}\chi ^{\beta }(\mathbf{r}^{\prime
})\right) ,\no
\end{eqnarray}%
where the frequency matrix $\Delta ^{j}$ was introduced, 
\[
\left( \Delta ^{j}\right) _{nm}=\delta _{j,n-m}, 
\]%
and $\alpha $ and $\beta $ stand for spin components.

The integrand in the interaction term, Eq. (\ref{Sint1}), contains a product
of two ``scalar products'' of the vectors $\chi $. Using the anticommutation
relations for the Grassmann variables $\chi $ we can also rewrite Eq. (\ref%
{Sint1}) in an equivalent form%
\begin{eqnarray}
S_{int} &=&\frac{T}{2}\int d\mathbf{r}d\mathbf{r}^{\prime }\left( \bar{\chi}%
^{\alpha }\left( \mathbf{r}\right) \Delta ^{j}\chi ^{\beta }\left( \mathbf{r}%
^{\prime }\right) \right)  \label{e2} \\
&&\times V_{0}\left( \mathbf{r-r}^{\prime }\right) \left( \bar{\chi}^{\beta
}\left( \mathbf{r}^{\prime }\right) \Delta ^{-j}\chi ^{\alpha }\left( 
\mathbf{r}\right) \right)  \nonumber
\end{eqnarray}

Now the crucial step is to guess an interaction term that could be written
in terms of the supervectors $\psi $. We have understood that for the
noninteracting part $S_{0}$, Eq. (\ref{eq:2.15}), we could simply substitute
the vectors $\chi $ by the supervectors $\psi $ and obtain $S_{0}^{\prime }$, Eq. (\ref{e4}). It is clear that just replacing the vectors $\chi $ by the
supervectors $\psi $ in Eq. (\ref{Sint1}) cannot be the resolution of the
problem, since in this way one takes into account  ``Fock''
diagrams in the first order but cannot obtain the ``Hartree'' ones. In contrast, replacing the
vectors $\chi $ by the supervectors $\psi $ in Eq. (\ref{e2}) one can
reproduce the ``Hartree'' type diagrams but the ``Fock'' diagrams are zero.

Actually, this property is the key to the construction of the new
supersymmetric model. In order to take into account both ``Hartree'' and
``Fock'' type diagrams we write the new interaction term $S_{int}^{\prime }$
in the following form%
\begin{eqnarray}
&&S_{int}^{\prime } =-\frac{T}{2}\int d\mathbf{r}d\mathbf{r}^{\prime }
\label{e6} \\
&&\;\;[\left( \bar{\psi}^{\alpha }\left( \mathbf{r}\right) \Delta ^{j}\psi
^{\alpha }\left( \mathbf{r}\right) \right) V_{0}\left( \mathbf{r-r}^{\prime
}\right) \bar{\psi}^{\beta }\left( \mathbf{r}^{\prime }\right) \Delta
^{-j}\psi ^{\beta }\left( \mathbf{r}^{\prime }\right)  \nonumber \\
&&\;\;-\left( \bar{\psi}^{\alpha }\left( \mathbf{r}\right) \Delta ^{j}\psi
^{\beta }\left( \mathbf{r}^{\prime }\right) \right) V_{0}\left( \mathbf{r-r}%
^{\prime }\right) \left( \bar{\psi}^{\beta }\left( \mathbf{r}^{\prime}\right) \Delta
^{-j}\psi ^{\alpha }\left( \mathbf{r}\right) \right) ]  \nonumber
\end{eqnarray}

The structure of the two terms in the integrand in Eq. (\ref{e6}) repeats
the structure of Eqs. (\ref{Sint1}) and (\ref{e2}). However, due to the
supervector structure of $\psi $ these terms are not equal to each other,
which contrast the case with the anticommuting vectors $\chi $. Moreover, if 
$\psi $ contained only bosonic variables these two terms would cancel each
other completely.

Equations (\ref{e4}) and (\ref{e6}) give the action 
\begin{equation}
S^{\prime }=S_{0}^{\prime }+S_{int}^{\prime }  \label{e7}
\end{equation}%
of the new supersymmetric model we want to use. The introduction of this
model is the main step toward the derivation of the supermatrix $\sigma $%
-model.

Before starting the derivation of the $\sigma $-model it is very important
to understand which important features of the original model are kept in the
supersymmetric one and what is lost. \ Of course, the partition function $%
Z^{\prime }$
\begin{equation}
Z^{\prime }=\int \exp \left( -S^{\prime }\right) D\psi  \label{e6a}
\end{equation}%
of the supersymmetric model is exactly equal to unity for any parameters of
the Hamiltonian, which is a consequence of the supersymmetry. Therefore, a
comparison of the partition functions of the models does not make sense.
At the same time, the partition function of the initial model does not
contain any singularities in any dimension and this is not a quantity we
want to study.

Interesting properties of the model can be seen in such quantities as
conductivity, tunnelling density of states, etc. These quantities are
expressed in terms of one- and two-particle Green functions. So, we compare
the initial electron model with the supersymmetric one by writing diagrams
describing lowest orders of the perturbation theory in interaction for Green
functions. When discussing one particle Green functions we compare the
self-energies of the Green functions, whereas vertices can be calculated for
two particle correlation functions. The following correlation functions are most interesting for us:

The imaginary time Green function,
\be
\mathcal{G}_{\sigma\sigma'}(\bfr,\bfrp,\tau,\tau')&=&-\left\langle \chi_\sigma(\bfr,\tau)\chi_{\sigma'}^*(\bfrp,\tau') \right\rangle_S\\
\left\langle\dots\right\rangle_S&=&\frac{\int D(\chi^*,\chi)\;(\dots)\;\mbox{e}^{-S}}{\int D(\chi^*,\chi)\;\mbox{e}^{-S}}.
\ee
It can be expressed as
\be
&&\mathcal{G}_{\sigma\sigma'}(\bfr,\bfrp,\tau,\tau')\no\\
&=&T^2\sum_{mn} \mathcal{G}_{\sigma\sigma'}\left( \mathbf{r,r}^{\prime },\omega_m,\omega_n\right)\;\mbox{e}^{-i\omega_m\tau+i\omega_n\tau'},
\ee
and calculated with the supersymmetric model
\be
\mathcal{G}_{\sigma\sigma'}\left( \mathbf{r,r}^{\prime },\omega_m,\omega_n\right) &=&\frac{i}{T}\left\langle \chi_\sigma \left( 
\mathbf{r},m\right) \bar{\chi}_{\sigma'}\left( \mathbf{r}^{\prime },n\right)\right\rangle_{S'}\no\\&&\label{e8}\\
\left\langle\dots\right\rangle_{S'}&=&\int D(\bar{\psi},\psi)\;(\dots)\;\mbox{e}^{-S'}
\label{e8a}.
\ee
The correlation function $\Pi(\bfr,\bfr',\tau,\tau')$.
\be
\Pi (\mathbf{r},\mathbf{r}^{\prime },\tau ,\tau ^{\prime })&=&\left\langle\!\left\langle
\rho (\mathbf{r},\tau )\rho (\bfr^{\prime },\tau ^{\prime })
\right\rangle\!\right\rangle_S \\
\rho (\mathbf{r},\tau )&=&\sum_{\sigma }\chi _{\sigma }^{\ast }(\mathbf{r}%
,\tau )\chi _{\sigma }(\mathbf{r},\tau )
\ee
By $\left\langle \!\left\langle \dots \right\rangle \!\right\rangle $ we
denote the connected part of the averages, $\left\langle AB\right\rangle
=\left\langle \!\left\langle AB\right\rangle \!\right\rangle +\left\langle
A\right\rangle \left\langle B\right\rangle $. We rewrite this correlation function in the form 
\be
\Pi(\bfr,\bfrp,\tau,\tau')&=&T^2\sum_{kl}\;\Pi(\bfr,\bfrp,\Omega_k,\Omega_l)\;\mbox{e}^{-i\Omega_k\tau+i\Omega_l\tau'}\no,\\&&\\
\Pi(\bfr,\bfrp,\Omega_k,\Omega_l)&=&-\left\langle \bar{\chi}(\bfr)\Delta^{-k}\chi(\bfr)\;\bar{\psi}(\bfrp)\Delta^{l}\psi(\bfrp)\right\rangle_{S'}.\no\\
\label{e9}
\ee
The average $\left\langle\dots\right\rangle_{S'}$ was defined in Eq.~(\ref{e8a}) above.

Equation (\ref{e8}) gives the one particle Green function for the supersymmetric
model, while Eq. (\ref{e9}) introduces the density-density correlation
function. In both cases the action $S^{\prime }$ is given by Eqs. (\ref{e4}) and (\ref{e6}). The presence of the variables $\chi $ in Eqs. (\ref{e8}) and (\ref{e9}) breaks the supersymmetry and
allows us to have nonzero results. Expanding the exponentials in $%
S_{int}^{\prime }$ and calculating the functional integrals over $\psi $ in
Eqs. (\ref{e8}) and (\ref{e9}) we obtain contributions that should be compared
with proper terms of the perturbation theory for the initial model.
Performing this calculation it is convenient to use a diagrammatic language,
which will be introduced now.

We have three building blocks:

\begin{description}
\item[(a)] The Green function of the noninteracting theory (exact in
disorder) is denoted by a solid line:

\includegraphics[width=0.3\linewidth]{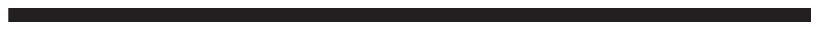}

\item[(b)] An interaction line corresponding to the first term in Eq. (\ref%
{e6}) (``Fock interaction'') is denoted by a single dotted line:

\includegraphics[width=0.3\linewidth]{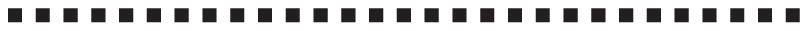}

\item[(c)] An interaction line corresponding to the second term in Eq. (\ref%
{e6}) (``Hartree interaction'') is denoted by a double dashed line:

\includegraphics[width=0.3\linewidth]{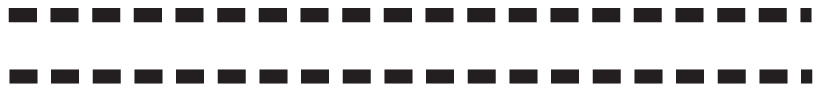}
\end{description}

Of course, both types of the interaction introduced in (\textbf{b}) and (%
\textbf{c}) correspond to the bare interaction $V_{0}\left( \mathbf{r-r}%
^{\prime }\right) $. The difference between them is related to how
they are connected to the superfield $\psi $.

\begin{figure}[tbp]
\begin{center}
\includegraphics[width=0.7\linewidth]{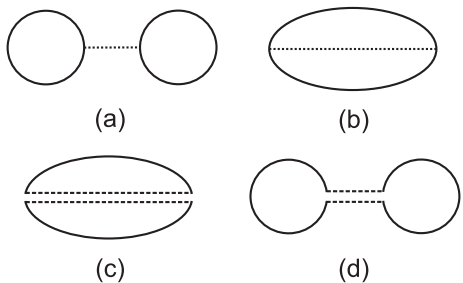}
\end{center}
\caption{The vacuum diagrams at first order in the interaction
potential. All of them vanish due to supersymmetry.}
\label{vacuum}
\end{figure}

We can check easily doing the perturbation theory that the partition
function $Z^{\prime }$ is normalized to unity. This property will be used in
the next section when performing the disorder averaging as for the
noninteracting case. The normalization is obvious for the noninteracting
part. On the other hand, for a perturbative expansion in $S_{int}^{\prime }$
the corresponding Gaussian integrals taken with $S_{0}^{\prime }$ vanish,
since the resulting expressions contain products of terms of the form $%
\mathrm{Str}(G^{n})=0$, where here $G$ denotes the average $\left\langle
\psi \bar{\psi}\right\rangle $ taken with $S_{0}^{\prime }$.
Diagrammatically, in the first order in the interaction one finds four
diagrams depicted in Fig.~\ref{vacuum}. 
Figures $(1b)$ and $(1d)$ can be written as a single supertrace, Figs. $%
(1a)$ and $(1c)$ as the product of two supertraces. Closed loops in the
diagrams correspond to supertraces, if we agree that each line of the
interaction (\textbf{c}) may be used to close a loop. Now we see why it is
convenient to use the single and double lines for the interaction. With this
convention all the four diagrams in Fig. \ref{vacuum} contain closed
``superparticle'' lines and vanish due to the supersymmetry. According to
our convention, Figs.~$\left( 1b\right) $ and $\left( 1d\right) $
contain one loop, while Figs.~$\left( 1a\right) $ and $\left( 1c\right) $ contain two
loops.

\begin{figure}[tbp]
\begin{center}
\includegraphics[width=0.9\linewidth]{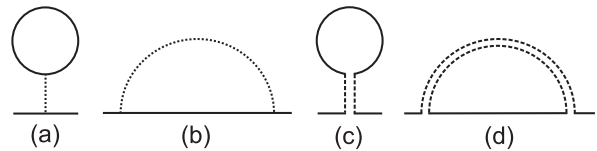}
\end{center}
\caption{Corrections to the Green function at first order in the
interaction potential. Diagrams (\textbf{a}) and (\textbf{d}) vanish,
diagrams (\textbf{b}) and (\textbf{c}) give the Hartree and Fock
contribution, respectively.}
\label{gcorrections}
\end{figure}

With this intuition we can discuss interaction corrections to Green
functions depicted in Fig.~{\ref{gcorrections}}. The first [Fig.~$\left( 2a\right) 
$] and fourth [Fig.~$\left( 2d\right) $] diagrams vanish, while the second [Fig.~$%
\left( 2b\right) $] and third [Fig.~$\left( 2c\right) $] give the Fock and Hartree
contributions. 
Note however, that screening of the interaction lines inside these diagrams
is not correctly reproduced. While the first [Fig.~$\left( 3a\right) $] and
second [Fig.~$\left( 3b\right) $] diagrams vanish, the
third [Fig.~$\left( 3c\right) $] diagram gives twice the desired contribution. The
reason is a simple combinatorial factor related to the fact that we work
with two interaction terms. 
\begin{figure}[tbp]
\begin{center}
\includegraphics[width=0.95\linewidth]{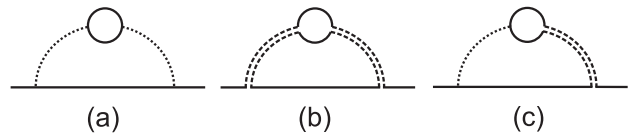}
\end{center}
\caption{Screening of the interaction lines. These are the diagrams at
second order in the interaction potential. The first and second diagrams are
equal to zero, the contribution of the third diagram is too large by a
factor of 2.}
\label{screening}
\end{figure}

This shows that our replacement of the initial electron model by the
supersymmetric one is not exact and its accuracy is restricted by the lowest
order in the interaction for the self-energy. The screening of the
interaction lines in the self-energy is described by higher orders in the
interaction and is not given correctly by the supersymmetric model.

\begin{figure}[tbp]
\includegraphics[width=0.99\linewidth]{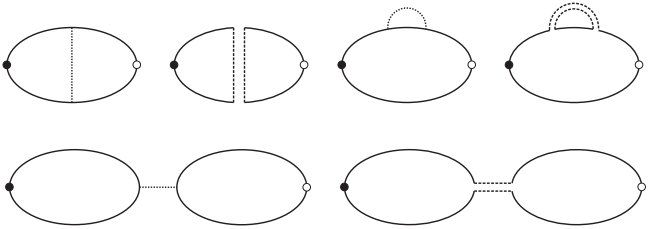} 
\caption{First order diagrams for the density-density correlation function.
The symbol $\bullet $ stands for a source term that distinguishes between
fermionic and bosonic variables, while $\circ $ symbolizes a supersymmetric
source term. Choosing the source terms in this way, no artificial
contributions are generated in this approximation.}
\label{correlation}
\end{figure}

At the same time, the density-density correlation function $\Pi $, Eq. (\ref%
{e9}), can also be correctly described by the supersymmetric model. In
addition to the corrections to the self-energy of the Green functions the
density-density correlation function contains vertex corrections represented
in Fig.~\ref{correlation}. In order to write correctly this function in
terms of the integral over the supervectors $\psi $ one should have both
supersymmetric and nonsupersymmetric vertices, which corresponds to the
definition, Eq. (\ref{e9}). Again, using our two types of interaction
lines we can check that the ladder diagrams for the density-density
correlation function are reproduced correctly. Diagrams containing closed
supersymmetric loops (the second and the fourth diagrams in the first line
and the first diagram in the second line) are equal to zero but the
remaining diagrams (e.g., the first and the third diagram in the first line
of Fig.~\ref{correlation} and the second diagram of the second line) do the
job properly. One can easily extend this to higher order ladder diagrams.

Thus, we see that the replacement of the initial electron model by the
supersymmetric one allows us to reproduce reasonably the main order in the
interaction. Let us emphasize that the consideration of the perturbation
theory in the interaction presented in this section was performed using
Green functions for a given disorder potential. No averaging over the
disorder or making approximations with respect to it was necessary.

At the same time, our aim is to derive a $\sigma $-model, which implies an
averaging over the disorder. The form of the supersymmetric $\sigma $-model
allows us to perform the averaging over the disorder in exactly the same way
as it has been done for the noninteracting case. The scheme of the
derivation is similar to the one used by Finkel'stein \cite%
{Finkelstein83,Finkelstein90} and we can simply follow his procedure.

\section{$\ \ \protect\sigma $-model with interaction}

\label{sec:smodel} Now we construct a low energy theory for the model
introduced in the preceding section. After the disorder averaging and the
decoupling of the quartic term by integration over the supermatrix field $Q(%
\mathbf{r})$ in complete analogy with what has been done for the
noninteracting case one finds 
\begin{eqnarray}
&&Z^{\prime }=\int D(\psi ,\bar{\psi})DQ\;\exp \left( -\frac{\pi \nu }{4\tau 
}\int d\mathbf{r}\;\mathrm{Str}\,Q^{2}(\mathbf{r})\right)  \no \\
&&\times \exp \left( i\int d\mathbf{r}\;\bar{\psi}(\mathbf{r})\left[ iE+\hat{%
\xi}_{p}+\frac{i}{2\tau }Q(\mathbf{r})\right] \psi (\mathbf{r})\right) 
\nonumber \\
&&\times \exp \left( -S_{int}^{\prime }[\psi ,\bar{\psi}]\right) .\label{e11}  
\end{eqnarray}%
where the interaction term $S_{int}^{\prime }$ is given by Eq. (\ref{e6}).

As in Ref.~\cite{Finkelstein83}, we should single out in the quartic
interaction $S_{int}^{\prime }\left[ \psi ,\bar{\psi}\right] $ pairs slowly
varying in space and time. However, in contrast to the replica approach, our
supersymmetric interaction, Eq. (\ref{e6}), contains two terms. Hence, after
singling out the slow pairs we obtain four terms instead of the two terms in 
Ref. \cite{Finkelstein83}. At the same time, one obtains as usual only two
interaction amplitudes $\Gamma _{1}$ and $\Gamma _{2}$ describing the small
and large angle scattering. These amplitudes can be obtained by the
replacement $V_{0}(\mathbf{q})\rightarrow \Gamma _{1}/\nu $ and $V_{0}(%
\mathbf{p}_{1}-\mathbf{p}_{2}-\mathbf{q})\rightarrow \Gamma _{2}/\nu $ for
small momenta $\bfq$. The quantities $\Gamma _{1}$ and $\Gamma _{2}$ allow to
make contact with the Fermi liquid theory. In particular, the singlet and
triplet amplitudes $\Gamma _{s}=\Gamma _{1}-\frac{1}{2}\Gamma _{2}$ and $%
\Gamma _{t}=\frac{1}{2}\Gamma _{2}$ are related to the Fermi-liquid
constants $F_{a}^{0}$ and $F_{s}^{0}$, 
\[
\Gamma _{s}=\frac{1}{2}\frac{F_{s}^{0}}{1+F_{s}^{0}},\quad \Gamma _{t}=-%
\frac{1}{2}\frac{F_{a}^{0}}{1+F_{a}^{0}}. 
\]

(We remind, however, that we may work only in the limit of small $\Gamma
_{1} $ and $\Gamma _{2}$).

When integrating over the fields $\psi ,\bar{\psi}$ we consider the simplest
possible approximation replacing $\exp \left( -S_{int}^{\prime }\right) $ by 
$\exp (-\left\langle S_{int}^{\prime }\right\rangle )$ for the interaction
term, where averaging is with respect to the quadratic form in $\psi $ in
Eq. (\ref{e11}). Using the relation $\left\langle \psi (\mathbf{r})\bar{\psi}%
(\mathbf{r^{\prime }})\right\rangle =g(\mathbf{r},\mathbf{r}^{\prime }),$
where $g\left( \mathbf{r,r}^{\prime }\right) $ is the Green function
corresponding to the quadratic in $\psi $ form of Eq. (\ref{e11}), we get 
\begin{widetext}
\be
\left\langle S_{int}'\right\rangle&=&-\frac{T}{2}\int (d\bfp_1) (d\bfp_2) (d\bfq) \;\Str\big[\Delta^jg^{\alpha\alpha}({\bf p}_1-{\bf q},{\bf p}_1)\big]\;V_0({\bf q})\;\Str\big[\Delta^{-j}g^{\beta\beta}({\bf p}_2+{\bf q},{\bf p}_2)\big]\no\\
&&+\frac{T}{2}\int (d\bfp_1) (d\bfp_2) (d\bfq) \;\Str\big[\Delta^jg^{\beta\alpha}({\bf p}_1-{\bf q},{\bf p}_1)\Delta^{-j}g^{\alpha\beta}({\bf p}_2+{\bf q},{\bf p}_2)\big]\;V_0({\bf p}_1-{\bf p}_2-{\bf q})\no\\
&&+\frac{T}{2}\int (d\bfp_1) (d\bfp_2) (d\bfq) \;\Str\big[\Delta^jg^{\beta\alpha}({\bf p}_1-{\bf q},{\bf p}_1)\big]\;V_0({\bf p}_1-{\bf p}_2-{\bf q})\;\Str\big[\Delta^{-j}g^{\alpha\beta}({\bf p}_2+{\bf q},{\bf p}_2)\big]\no\\
&&-\frac{T}{2}\int (d\bfp_1) (d\bfp_2) (d\bfq) \;\Str\big[\Delta^jg^{\alpha\alpha}({\bf p}_1-{\bf q},{\bf p}_1)\Delta^{-j}g^{\beta\beta}({\bf p}_2+{\bf q},{\bf p}_2)\big]\;V_0({\bf q}).
\ee
\end{widetext}From the momentum flow in the interaction potential one reads
off processes with small and large angle scattering, which should be
described by amplitudes $\Gamma _{1}$ and $\Gamma _{2}$, respectively. The
singling out of the slow modes corresponds to a restriction of small momenta 
$\mathbf{q}$ and frequencies. Then, calculating the integral over $Q$ we may
use the saddle point approximation that gives for the saddle point $g(%
\mathbf{r},\mathbf{r})=\pi \nu Q(\mathbf{r})$. Then, the effective free
energy can be written as%
\begin{eqnarray}
&&F_{int}=\left\langle S_{int}^{\prime }\right\rangle  \nonumber \\
&=&-\frac{(\pi \nu )^{2}T}{2}\int d\mathbf{r}\;\Gamma _{1}\;\left\{ \mathrm{%
Str}\big[\Delta ^{j}Q^{\alpha \alpha }\big]\;\mathrm{Str}\big[\Delta
^{-j}Q^{\beta \beta }\big]\right.  \nonumber \\
&&\qquad \qquad \left. +\mathrm{Str}\big[\Delta ^{j}Q^{\alpha \alpha }\Delta
^{-j}Q^{\beta \beta }\big]\right\}  \nonumber \\
&&+\frac{(\pi \nu )^{2}T}{2}\int d\mathbf{r}\;\Gamma _{2}\;\left\{ \mathrm{%
Str}\big[\Delta ^{j}Q^{\beta \alpha }\big]\;\mathrm{Str}\big[\Delta
^{-j}Q^{\alpha \beta }\big]\right.  \nonumber \\
&&\qquad \qquad \left. +\mathrm{Str}\big[\Delta ^{j}Q^{\beta \alpha }\Delta
^{-j}Q^{\alpha \beta }\big]\right\} .
\end{eqnarray}%
Working with the saddle point of the noninteracting theory is a frequently
used approximation in the context of $\sigma$-models for interacting disordered
systems \cite{Finkelstein90,Belitz94}. The derivation for the remaining
parts of the free energy can be performed exactly in the same way as in the
replica approach. As a result, we come to the free energy functional $F\left[
Q\right] $ describing the supermatrix $\sigma $-model, 
\begin{eqnarray}
F\left[ Q\right] &=&\frac{\pi \nu }{4}\int d\mathbf{r}\;\mathrm{Str}\big[%
D(\nabla Q)^{2}-4EQ]  \nonumber \\
&+&\frac{\pi \nu }{4}\int d\mathbf{r}\;\big[\Gamma _{2}Q\gamma _{2}Q-\Gamma
_{1}Q\gamma _{1}Q\big]  \label{eq:ssigmamodel1}
\end{eqnarray}%
where 
\begin{eqnarray}
Q\gamma _{1}Q &=&2\pi T\sum_{n_{i}\alpha \beta }\left( \mathrm{Str}\big[%
Q_{n_{1}n_{2}}^{\alpha \alpha }\big]\mathrm{Str}\big[Q_{n_{3}n_{4}}^{\beta
\beta }\big]\right.  \nonumber \\
&&\qquad \left. +\mathrm{Str}\big[Q_{n_{1}n_{2}}^{\alpha \alpha
}Q_{n_{3}n_{4}}^{\beta \beta }\big]\right) \delta
_{n_{1}+n_{3},n_{2}+n_{4}},\quad  \label{eq:gamma1} \\
Q\gamma _{2}Q &=&2\pi T\sum_{n_{i}\alpha \beta }\left( \mathrm{Str}\big[%
Q_{n_{1}n_{2}}^{\alpha \beta }\big]\mathrm{Str}\big[Q_{n_{3}n_{4}}^{\beta
\alpha }\big]\right.  \nonumber \\
&&\qquad \left. +\mathrm{Str}\big[Q_{n_{1}n_{2}}^{\alpha \beta
}Q_{n_{3}n_{4}}^{\beta \alpha }\big]\right) \delta _{n_{1}+n_{3},n_{2}+n_{4}}.
\label{eq:gamma2}
\end{eqnarray}%
Again, we note the constraints $Q^{2}(\mathbf{r})=1$ and $Q^{\dagger }(%
\mathbf{r})=KQ(\mathbf{r})K$, where $K$ is the matrix 
\[
K=\left( 
\begin{array}{cc}
1 & 0 \\ 
0 & k%
\end{array}%
\right) _{AR},\qquad k=\left( 
\begin{array}{cc}
1 & 0 \\ 
0 & -1%
\end{array}%
\right) _{BF}. 
\]%
Matrices $K$ and $k$ act in the advanced/retarded ($AR$) and boson/fermion ($%
BF$) sectors, respectively.

The interaction part of the supermatrix $\sigma $-model, Eqs. (\ref%
{eq:gamma1}) and (\ref{eq:gamma2}) contains, in contrast to the replica $\sigma $%
-model, in total four terms and is somewhat more complicated. At the same
time, its form is quite similar to the one in the replica approach. It is
important to emphasize that the $\sigma $-model obtained is fully
supersymmetric as it should be. The supersymmetry can be violated by source
terms that can be added for generating correlations functions.

For actual calculations it can be convenient to express the interaction part
in a different way, using auxiliary fields $\varphi _{i},\phi _{i}$. 
\begin{widetext}
\be
\label{eq:fintwithphi}
F_{int}&=&-\frac{(\pi\nu)^2}{2}\int d\bfr d\bfrp\;\left(\left\langle\Str\big[\underline{\varphi}_1(\bfr)Q(\bfr)\big]\Str\big[\underline{\varphi}_1(\bfrp) Q(\bfrp)\big]\right\rangle_{\varphi_1}+\left\langle\Str\big[\underline{\phi}_1(\bfr)Q(\bfr)\big]\Str\big[\underline{\phi}_1(\bfrp)Q(\bfrp)\big]\right\rangle_{\phi_1}\right)\no\\
&&+\frac{(\pi\nu)^2}{2}\int d\bfr d\bfrp\;\left(\left\langle\Str\big[\underline{\varphi}_2(\bfr)Q(\bfr)\big]\Str\big[\underline{\varphi}_2 (\bfrp) Q(\bfrp)\big]\right\rangle_{\varphi_2}+\left\langle\Str\big[\underline{\phi}_2(\bfr)Q(\bfr)\big]\Str\big[\underline{\phi}_2(\bfrp)Q(\bfrp)\big]\right\rangle_{\phi_2}\right).
\ee
\end{widetext}In this way one can work with expressions including only
supertraces and not single components. This makes the structure more
transparent and additionally it has technical advantages when working with
supermatrices, since anticommutativity of the fermionic components is taken
care of automatically. The fields $\varphi _{i},\phi _{i}$ do not have a
nontrivial structure in the $AR$ space. Instead, this structure is given by 
\[
\underline{\phi _{i}}(\mathbf{r})=T\sum_{j}\phi _{i}(\mathbf{r},j)\Delta
^{j}, 
\]%
where $j$ is a frequency index. The same definition is used for $\varphi
_{i} $. The fields $\varphi _{1},\varphi _{2}$ have only commuting entries
and are proportional to the unit matrix, while $\phi _{1},\phi _{2}$ are
conventional supermatrices. For convenience, the explicit form together with
the Gaussian integrals used to define the averaging are given in the
appendix \ref{appsec:parameterization}. Of particular importance are
relations giving the averages of the fields 
\begin{eqnarray}
&&\Big\langle\varphi _{1}(\mathbf{r},j)\varphi _{1}(\mathbf{r^{\prime }}%
,-j)+\phi _{1}(\mathbf{r},j)\phi _{1}(\mathbf{r^{\prime }},-j)\Big\rangle%
_{\mu \eta }^{\alpha \delta }  \nonumber \\
&=&\delta _{\mu \eta }^{\alpha \delta }\;\;\;\frac{\Gamma _{1}}{\nu T}%
\;\delta (\mathbf{r}-\mathbf{r^{\prime }})  \label{unity1}
\end{eqnarray}%
and 
\begin{eqnarray}
&&\Big\langle\varphi _{2}(\mathbf{r},j)\varphi _{2}(\mathbf{r^{\prime }}%
,-j)+\phi _{2}(\mathbf{r},j)\phi _{2}(\mathbf{r^{\prime }},-j)\Big\rangle%
_{\mu \eta }^{\alpha \delta }  \nonumber \\
&=&2\;\delta _{\mu \eta }^{\alpha \delta }\;\;\;\frac{\Gamma _{2}}{\nu T}%
\;\delta (\mathbf{r}-\mathbf{r^{\prime }}),  \label{unity2}
\end{eqnarray}%
where upper indices refer to spin, lower indices to fermion-boson space. The
importance of these relations will become obvious later.

Using the formalism developed in this section we can express the quantities
like Green function in coinciding points determining the
tunnelling density of states or the density-density correlation function in
terms of a functional integral over the supermatrices $Q$ with the free
energy functional $F$, Eq. (\ref{eq:ssigmamodel1}). This can be done using Eq.
(\ref{e11}) and integrating over $\psi $ in Eqs. (\ref{e8}) and (\ref{e9}) with
the effective free energy functional of Eq. (\ref{e11}) containing both $Q$
and $\psi $.
The following simple formulas are obtained when averaging the interaction part and the pre-exponential factors in Eqs.~(\ref{e8}) and (\ref{e9}) separately. Furthermore the saddle point equation $Q(\bfr)=\pi\nu g(\bfr,\bfr)$ (\ref{eq:sp}) is used,  
\be
\mathcal{G}_{\sigma\sigma'}\left( \mathbf{r,r},\omega_m,\omega_n\right)=\frac{i\pi\nu}{T}\int DQ \;\Str\big[k_+Q_{mn}^{\sigma\sigma'}(\bfr)]\;\mbox{e}^{-F}.\no\\ \label{e17}
\ee
For the density-density correlation we consider only one out of the two possible pairings for the pre-exponential factors in Eq.~(\ref{e8}), since we are interested in points $\bfr,\bfrp$ that are far apart and $g(\bfr,\bfrp)$ falls off at distances $|\bfr-\bfrp|$ of the order of the mean free path,
\be
&&\Pi \left(\mathbf{r,r}^{\prime},\Omega_k,\Omega_l\right) \label{e18}\\
&=& -(\pi\nu)^2 \int DQ\;\Str\big[k_+\Delta^{-k}Q(\bfr)\big]\Str\big[\Delta^{l}Q(\bfrp)\big]\mbox{e}^{-F}.\no
\ee
Equations (\ref{eq:ssigmamodel1}), (\ref{eq:gamma1}), (\ref{eq:gamma2}), (\ref{e17}), and 
(\ref{e18}) are the final results of the derivation of the $\sigma $-model
with interaction. Any physical quantity we are interested in can be computed,
at least in principle, using these formulas. Of course, our ultimate goal is
to compute some new quantities in the nonperturbative regime. However, in
order to see how the model works we should understand first how it can help
to reproduce known perturbative results. Such calculations will be carried
out in the next sections.

\section{Perturbation theory in diffusion modes}

\label{sec:perturbationtheory}Disordered systems with interaction can be
studied using a conventional diagrammatic technique (see, e.g., Ref.~\cite%
{Altshuler85}). In low dimensions, the most important contributions come
from a certain class of diagrams called ``cooperons'' and ``diffusons''.
Although these diagrams are important even without any interaction, the
latter leads to new contributions. The $\sigma $-model formalism allows one
to integrate over ``electron degrees of freedom'' in the beginning of all
calculations and reduce them to study of the low lying diffusion modes. In
the language of the $\sigma $-model summation of the diffusons and cooperons
is equivalent to an expansion in small fluctuations near the ground state of
the free energy functional $F$, Eq. (\ref{eq:ssigmamodel1}) . The minimum of 
$F$ is achieved setting $Q=\Lambda $, where $\Lambda $ is defined in Eq. (%
\ref{e12}).

For a perturbative expansion the matrix $Q$ may be parametrized in the
vicinity of $\Lambda $ as 
\begin{equation}
Q=\Lambda (1+iP)(1-iP)^{-1},  \label{e14}
\end{equation}%
where $P=KP^{\dagger }K$, 
\begin{eqnarray}
P &=&\left( 
\begin{array}{cc}
0 & B \\ 
kB^{\dagger } & 0%
\end{array}%
\right), \\
B &=&\left( 
\begin{array}{cc}
a & i\sigma \\ 
\rho ^{\dagger } & ib%
\end{array}%
\right) ,\qquad kB^{\dagger }=\left( 
\begin{array}{cc}
a^{\dagger } & \rho \\ 
i\sigma ^{\dagger } & ib^{\dagger }%
\end{array}%
\right).
\end{eqnarray}%
Following the structure of $\psi $ all matrices in $AR$-space are now
arranged in block form as 
\[
\left( 
\begin{array}{cc}
\left. M(k,l)\right| _{k<0,l<0} & \left. M(k,l)\right| _{k<0,l\geq 0} \\ 
\left. M(k,l)\right| _{k\geq 0,l<0} & \left. M(k,l)\right| _{k\geq 0,l\geq 0}%
\end{array}%
\right) . 
\]%
The perturbation theory in the diffusion modes is carried out by making
expansion in the supermatrices $P$. Making expansion in $P$ we represent
first the free energy functional $F$ in a form of a series  in $P$. As the
zero approximation we take the noninteracting quadratic part $F_{0}$ 
\begin{equation}
F_{0}=\pi\nu\int d\bfr\;\Str\big[D\nabla P(\bfr)\nabla P(\bfr)+2E\Lambda P^2(\bfr)\big].
\label{e15}
\end{equation}%
The rest of the functional consists of quadratic terms coming from the
interaction, Eqs.(\ref{eq:gamma1}) and (\ref{eq:gamma2}), and higher order in $P$
terms coming from both noninteracting and interacting parts of the
functional $F$. For explicit calculations we must calculate averages of
different powers of $P$ with the bare functional $F_{0}$, Eq. (\ref{e15}).
Calculating Gaussian integrals it is not difficult to derive the following
contraction rules for arbitrary supermatrices $M,N$ (we remind the reader
that the unitary case is considered) 
\begin{eqnarray}
&&\big\langle\mathrm{Str}\left[ M_{k^{\prime }l^{\prime }}P_{lk}(\mathbf{r})%
\right] \;\mathrm{Str}\left[ N_{i^{\prime }j^{\prime }}P_{ji}(\mathbf{%
r^{\prime }})\right] \big\rangle_{0}  \nonumber \\
&=&\mathrm{Str}\left[ M_{k^{\prime }l^{\prime }}^{\perp }N_{i^{\prime
}j^{\prime }}^{\perp }\right] \frac{1}{2\pi \nu }\mathcal{D}\left( \mathbf{r}%
-\mathbf{r^{\prime }},k-l\right)  \nonumber \\
&&\quad \times \delta _{jk}\delta _{il}\Theta (l,l^{\prime })\Theta
(k,k^{\prime })  \label{contract1}
\end{eqnarray}%
and 
\begin{eqnarray}
&&\big\langle\mathrm{Str}\left[ M_{i^{\prime }j^{\prime }}\,P_{jk}(\mathbf{r}%
)\,N_{k^{\prime }l^{\prime }}\,P_{li}(\mathbf{r^{\prime }})\right] %
\big\rangle_{0}  \nonumber \\
&=&\Big[\mathrm{Str}M_{i^{\prime }j^{\prime }}\;\,\mathrm{Str}N_{k^{\prime
}l^{\prime }}-\mathrm{Str}M_{i^{\prime }j^{\prime }}\Lambda \;\,\mathrm{Str}%
N_{k^{\prime }l^{\prime }}\Lambda \Big]  \nonumber \\
&&\times \frac{1}{4\pi \nu }\mathcal{D}\left( \mathbf{r}-\mathbf{r^{\prime }}%
,j-k\right) \delta _{kl}\delta _{ji}\Theta (k,k^{\prime })\Theta
(j,j^{\prime }).\quad  \label{contract2}
\end{eqnarray}%
The subscript ``$0$'' will be omitted in the following. We denote $%
M^{\perp }=(M-\Lambda M\Lambda )/2$, and $M^{\parallel }=(M+\Lambda M\Lambda
)/2$. The diffusion propagator $\mathcal{D}$ is given by 
\begin{equation}
\mathcal{D}(\mathbf{r}-\mathbf{r^{\prime }},l-k)=\int (d\mathbf{q})\;\frac{%
\mbox{e}^{i\mathbf{q}(\mathbf{r}-\mathbf{r^{\prime }})}}{D\mathbf{q}%
^{2}+|\Omega _{l-k}|}.  \label{e16}
\end{equation}

and the symbol $\Theta (k,l)$ stands for 
\[
\Theta (k,l)=\left\{ 
\begin{array}{cc}
1 & \text{for }k\geq 0,\text{ }l\geq0\text{ \ \ or \ }k<0,\text{ \ }l< 0
\\ 
0 & \text{otherwise}%
\end{array}%
\right. 
\]

Correlation functions can be calculated expanding Eqs.~(\ref{e17}) and (\ref{e18})
) in $P$ and integrating over this variable with the help of the contraction
rules, Eqs. (\ref{contract1}) and (\ref{contract2}). A few elementary diagrams
for the density-density correlation function are depicted in Fig.~{\ref%
{correlation}}. As discussed in Sec.~\ref{sec:interactionpart}  one can read
off from the diagrams in Fig.~{\ref{correlation}} that only \emph{one} vertex term should distinguish between the
fermionic and the bosonic sector, i.e., contain only fermionic variables, while the other remains supersymmetric. In this way no artificial
contributions arise from the diagrams shown in Fig.~{\ref{correlation}. The same argument can be generalized straightforwardly to the ladder-type diagrams considered below. Using two vertex terms of the same kind would not lead to correct results. In this case either all diagrams vanish identically if both vertex terms are supersymmetric, or all diagrams in Fig.~{\ref{correlation}} give a contribution, if both vertex terms contain fermionic variables only.

Calculating the density-density correlation function $\Pi(\bfq,\Omega)$ we are interested in
small frequencies and momenta. In the static limit it is related to the compressibility $\partial_\mu n$ according to
\be
\partial_\mu n=\lim_{q\rightarrow 0} \Pi(\bfq,0).\no
\ee
For systems without interaction the relation $\partial_\mu n=2\nu$ holds. One can calculate $\Pi \left( \mathbf{q,}\Omega
_{k}\right) $ from the supersymmetric model representing this function as 
\begin{eqnarray}
&&\Pi (\mathbf{q},\Omega _{k})=\frac{\partial n}{\partial \mu }+\frac{1}{%
(1+F_{s}^{0})^{2}}\hat{\Pi}(\mathbf{q},\Omega _{k})  \label{pi}, \\
&&\left\langle F_{\varphi }F_{\vartheta }\right\rangle \sim T\sum_{k}\int_{q}\;\varphi (-\mathbf{q},-k)\hat{\Pi}(\mathbf{q},\Omega
_{k})\vartheta (\mathbf{q},k) \;.\quad
\label{pihat}
\end{eqnarray}%
The terms $F_{\vartheta }$ and $F_{\varphi }$ are given by 
\begin{eqnarray}
&&F_{\vartheta }=i\pi \nu \;\int d\mathbf{r}\;\mathrm{Str}\big[\underline{%
\vartheta }Q\big] , \label{e21} \\
&&F_{\varphi }=i\pi \nu \;\int d\mathbf{r}\;\mathrm{Str}\big[k_{+}\underline{%
\varphi }Q\big],  \label{e22}
\end{eqnarray}%
where $k_{+}=(1+k)/2$. The factor $1/(1+F_{s}^{0})^{2}$ is due to vertex
corrections taken from Fermi-liquid theory. It is related to the amplitudes $%
\Gamma _{i}$ as $1/(1+F_{s}^{0})=1-2\Gamma _{1}+\Gamma _{2}$ and also
determines $\partial _{\mu }n=2\nu /(1+F_{s}^{0})$. We consider the
Fermi-liquid approximation \cite{Finkelstein90}, for which the calculation
amounts to the ladder summation depicted in Fig.~\ref{laddersummation}. In
the zero order one should expand $Q\sim \Lambda +2i\Lambda P$ in the
interaction terms, Eqs. (\ref{eq:gamma1}) and (\ref{eq:gamma2}), in the free
energy $F$ and in the pre-exponential, Eq. (\ref{pihat}). In both the cases
the saddle point $\Lambda $ does not contribute. 
\begin{figure}[tbp]
\begin{center}
\includegraphics[width=0.9\linewidth]{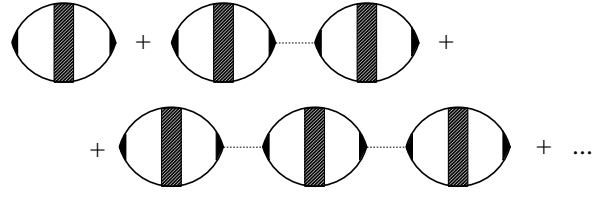}
\end{center}
\caption{Ladder summation for the density-density correlation function as
discussed in the text.}
\label{laddersummation}
\end{figure}
Only the singlet interaction amplitude $\Gamma _{s}$ enters and one must
sum a geometric series that arises after expanding in the interaction terms
written in the quadratic approximation in $P$. As a result, we obtain 
\[
\hat{\Pi}(\mathbf{q},\Omega _{k})=-2\nu \;\frac{|\Omega _{k}|}{D\mathbf{q}%
^{2}+z_{1}^{(0)}|\Omega _{k}|}, 
\]%
where $z_{1}^{(0)}=1+2\Gamma _{1}-\Gamma _{2}$. Using this relation and Eq. (%
\ref{pi}) one comes to the result 
\[
\Pi (\mathbf{q},\Omega _{k})=\partial _{\mu }n\frac{D_{FL}\bfq^2}{D_{FL}\mathbf{q}%
^{2}+|\Omega _{k}|}, 
\]%
where $D_{FL}=(1+F_{s}^{0})D$. The diffusion coefficient, which enters the
$\sigma$-model, can be related to conductivity with the help of the equation of
continuity 
\be
\sigma =e^{2}\;\lim_{\Omega \rightarrow 0}\lim_{p\rightarrow 0}\frac{\Omega 
}{\mathbf{p}^{2}}\;\Pi (\mathbf{p},\Omega ). \label{eq:continuity}
\ee
This leads to the Einstein relation 
\[
\sigma=e^{2}\partial _{\mu }nD_{FL}. 
\]%
This relation can be seen to hold also after renormalization\cite{Finkelstein83} (see Sec.~\ref{sec:rengroup} below) and in this way interaction corrections to conductivity can be obtained.

In the remaining part of this section, in contrast, we describe simple perturbation theory for finding the first order interaction correction to conductivity from the $\sigma$-model.  
A general expression for the calculation of the linear response conductivity can be given with the help of the original fermionic partition function $Z$ in Eq.~(\ref{eq:partition}),
\be
\frac{\sigma(\omega)}{e^2}&=&\left.\frac{K(\bfq\rightarrow 0,\Omega_n)}{\Omega_n}\right|_{i\Omega_n\to\omega^+},\\
K(\bfr,\tau,\bfrp,\tau')&=&\left.\left(-\frac{\delta^2}{\delta
A_\mu(\bfr,\tau)\delta A_\mu(\bfr',\tau')}\frac{Z[A]}{Z[0]}\right)\right|_{A=0},\no\\
&&\\
K(\bfq,\Omega_n)&=&\int_{\bfr,\tau}K(\bfr,\tau,0,0)
\;\mbox{e}^{-i\bfq\bfr}\;\mbox{e}^{i\Omega_n\tau}.
\ee
Analytic continuation is performed from positive Matsubara frequencies. 
No summation over $\mu$ is implied. One comes to $Z[A]$ after the substitution 
$\hat{p}\rightarrow \hat{p}-A$ in $Z=Z[0]$. The electron charge $e$ appears in the prefactor. 
For the $\sigma$-model the corresponding replacement 
is $\nabla Q\rightarrow \nabla Q-i[\underline{A},Q]$ leading to a change in the free energy 
$F=F[0]\rightarrow F[A]$. $A$ is chosen supersymmetric because we should treat fermions and bosons 
on equal footing for the derivation of the $\sigma$-model. In order to obtain the analog of $K$, however, 
it is necessary to break supersymmetry. To this end one should expand $Z'[A]=\int DQ \;\mbox{e}^{-F[A]}$ up 
to second order in $A$ and multiply one of the fields $A$ by the projector onto the fermionic sector $k_+$. 
Since we are interested in the leading order correction in the interaction we also expand up to first order 
in $F_{int}$. 
In order to express the result of this procedure in a compact form we define
\be
F_A^{(1)}&=&-\frac{\pi\nu i D}{2}\int d\bfr\;\Str\big[\nabla
Q(\bfr)[\underline{A}(\bfr),Q(\bfr)]\big],\\
F_A^{(1)+}&=&-\frac{\pi\nu i D}{2}\int d\bfr\;\Str\big[\nabla
Q(\bfr)[k_+\underline{A}(\bfr),Q(\bfr)]\big],\\
F_A^{(2)+}&=&-\frac{\pi\nu
D}{4}\int d\bfr\;\Str\big[[k_+\underline{A}(\bfr),Q(\bfr)][\underline{A}(\bfr),Q(\bfr)]\big].\no\\
\ee
We can now write the analog of $K$ for the supersymmetric model,
\be
&&K'(\bfr,\tau,\bfrp,\tau')=-\frac{\delta^2}{\delta
A_\mu(\bfr,\tau)\delta A_\mu(\bfrp,\tau')} \times\no\\
&&\times\left.\Big(-\left\langle F_{A}^{(2)+}\right\rangle_0+\frac{1}{2}\left\langle F_{A}^{(1)+}F_A^{(1)}\right\rangle_0\right.\no\\
&&\qquad\left.+\left\langle F_{A}^{(2)+}F_{int}\right\rangle_0-\frac{1}{2}\left\langle F_{A}^{(1)+}F_A^{(1)}F_{int}\right\rangle_0 \Big)\right|_{A=0}\no\\
&&=\sum_{i=1}^{4}K_i(\bfr,\tau,\bfrp,\tau'). \ee
Averaging is with respect to the noninteracting part of the free energy. We expand the matrix $Q$ in terms of $P$ and use $F_0$ (\ref{e15}) for averaging. Classical conductivity can be found at the level of the saddle point $Q=\Lambda$. The only contribution comes from $K_1'$ and one finds $\sigma_0=2e^2\nu D$, where the factor of $2$ is due to the spin degree of freedom.
Next we turn to interaction corrections. At order $P^4$ $K_4'$ vanishes in the limit $\bfq\rightarrow 0$. 
There are two contributions to $K_3'$,
\be
&&K_3'(\bfr,\bfrp,\tau,\tau')\\
&\sim&2(\pi\nu)^2 D\partial_{A_\mu}\partial_{A_\mu'} \int_{\bfr_i} \Big\langle \Str\big[k_+[\underline{A}_{\bfr_1},\Lambda P_{\bfr_1}][\underline{A}_{\bfr_1},\Lambda P_{\bfr_1}]\big] \no\\
&&\quad\times\sum_{j=1,2}(-1)^{j}\left\langle\Str\big[\underline{\hat{\phi}}_{j,\bfr_2}\Lambda P_{\bfr_2}\big]\Str\big[\underline{\hat{\phi}}_{j,\bfr_3}\Lambda P_{\bfr_3}\big]\right\rangle_{\hat{\phi}}\Big\rangle_0\no\\
&+&2(\pi\nu)^2 D\partial_{A_\mu}\partial_{A_\mu'} \int_{\bfr_i} \left\langle \Str\big[k_+[\underline{A}_{\bfr_1},\Lambda ][\underline{A}_{\bfr_1},\Lambda P^2_{\bfr_1}]\big] \right.\no\\
&&\left.\quad\times\sum_{j=1,2}(-1)^{j}\left\langle\Str\big[\underline{\hat{\phi}}_{j,\bfr_2}\Lambda P_{\bfr_2}\big]\Str\big[\underline{\hat{\phi}}_{j,\bfr_3}\Lambda P_{\bfr_3}\big]\right\rangle_{\hat{\phi}}\right\rangle_0.\no
\ee
After further evaluation one finds a cancellation and thus there is no correction including only two diffusons.
Considering now corrections with three diffusion modes the only nonvanishing contributions in the limit 
$\bfq\rightarrow 0$ can be seen to originate from $K_4'$, where all $Q$ matrices should be replaced by 
$2i\Lambda P$. At order $P^6$ we therefore have
\be 
&&{K'_4}^{(1)}(\bfr,\tau,\bfr',\tau')=4(\pi\nu)^4 D^2\partial_{A_\mu
A'_\mu}\\
&&\left\langle\int_{\bfr_i}\Str\Lambda\nabla
 P_{\bfr_1}[k_+\underline{A}_{\bfr_1},\Lambda P_{\bfr_1}]\;\Str \Lambda\nabla
P_{\bfr_2}[\underline{A}_{\bfr_2},\Lambda P_{\bfr_2}]\right.\no\\
&&\left.\sum_{j=1,2}(-1)^{j}\left\langle\Str\big[\underline{\hat{\phi}}_{j\bfr_3}\Lambda P_{\bfr_3}\big]\Str\big[\underline{\hat{\phi}}_{j,\bfr_4}\Lambda P_{\bfr_4}\big]\right\rangle\right\rangle_0\no\\
&=&2\langle\dots\rangle_{(\nabla1\nabla2)(13)(24)}+4\langle\dots\rangle_{(\nabla1
2)(13)(\nabla24)}\label{eq:3contractions}\\
&&+2\langle\dots\rangle_{(\nabla 1 3)(12)(\nabla24)}\no 
\ee
In the last two lines (\ref{eq:3contractions}) the relevant contractions and their multiplicities have 
been written. $(\nabla 1 3)$ symbolizes a contraction of $\nabla P(\bfr_1)$ and $P(\bfr_3)$, etc.
After taking the limit $\bfq\rightarrow 0$ one finds
\be
&&{K'_4}(\bfq\rightarrow 0,\Omega_n>0)
\\&=&-\frac{16D}{d}(\Gamma_1-2\Gamma_2)\int (d\bfp) \;D\bfp^2 \times\no\\ && T\left[\sum_{m=1}^{n-1}\;\Omega_m\;\mathcal{D}_{\bfp,m+n}\mathcal{D}^2_{\bfp,m}+\sum_{m\ge n}\;\Omega_n\;\mathcal{D}_{\bfp,m+n}\mathcal{D}^2_{\bfp,m}\right].\no
\ee
Analytic continuation for the conductivity gives
\be
\delta\sigma(\omega\rightarrow 0)&=&\frac{2i\sigma_0}{\pi d\nu}(\Gamma_1-2\Gamma_2)\int_{-\infty}^{\infty}dz\;\frac{\partial}{\partial z}\left(z \coth\frac{\beta z}{2}\right)\no\\
&&\times\int (d\bfq)\;\frac{D\bfq^2}{(D\bfq^2-iz)^3}.
\ee
In particular, for two dimensions and after replacing $\Gamma_i\rightarrow V_i$ one comes to the well 
known result\cite%
{Altshuler85}%
\begin{equation}
\delta \sigma =\sigma -\sigma _{0}=e^{2}/(2\pi^2 )(V_{1}-2V_{2})\ln (T\tau ).
\label{e25}
\end{equation}

This agreement demonstrates that the supersymmetric $\sigma $-model we have
derived reproduces correctly the known quantum corrections for a weak
interaction. However, this calculation cannot be considered as the crucial
check of the consistency of the $\sigma $-model because the first order
correction is not very sensitive to the details of the structure of the $%
\sigma $-model. Therefore, in the next section we proceed further
demonstrating the renormalizability of the model and deriving
renormalization group equations.

\section{Renormalization group}

\label{sec:rengroup}

The renormalization group (RG) procedure in the form introduced by
Finkel'stein \cite{Finkelstein90} will now be applied to the supersymmetric
model. Finkel'stein considered the problem of an interacting electron gas in
the diffusive regime. Based on the replica approach he developed an
appropriate $\sigma $-model description and demonstrated the
renormalizability up to the first order in $t=1/(4\pi ^{2}\nu D)$ and in all
orders in the interaction amplitudes. The scaling behavior was studied while
lowering the temperature of the system. It turned out to be possible to make
contact to the Fermi-liquid theory.

Typically, the (Fermi-liquid) interaction amplitudes are not small and in
the process of renormalization they can even diverge \cite%
{Finkelstein83,Castellani83}. In the original work \cite{Finkelstein83} the
case of long range Coulomb interaction was considered and the Cooperon modes
were suppressed (e.g., by a small magnetic field). Appropriate scaling
equations for this case were derived. Later on the approach has been
generalized to many different situations (for a review see Refs.~\cite%
{Finkelstein90} and \cite{Belitz94}). By a comparison to the field theoretic solution
the scaling equations can also be written starting from conventional
perturbation theory \cite{Castellani83} and for some calculations the
diagrammatics can also be conveniently used \cite{Aleiner99}(for a review
see, e.g., Ref.~\cite{DiCastro03}). The authors of Ref.~\cite{Castellani83} also
considered the case of short range interactions. A recent discussion in the
context of a Keldysh $\sigma $-model can be found in Ref.~\cite{Chamon99}.

The renormalization scheme for the interacting disordered electron gas is
more complicated than for the noninteracting case. ``Infrared''
singularities arise in the theory due to both the frequency and momentum
integration. It turns out that in addition to the diffusion coefficient and
the interaction amplitudes a ``charge'' $z$ multiplying the energy matrix
must be introduced. It describes the relative change of momentum and
frequency scales. Our aim is to show that the supersymmetric model
introduced above is capable of reproducing the correct RG equations in the
limit of weak short range interactions, i.e., in the first order in $t$ and
to leading order in the interaction amplitudes. Therefore the procedure is
used as a test for the new model and it is the scheme introduced by
Finkel'stein that serves this purpose best. As a by-product, the
Altshuler-Aronov corrections \cite{Altshuler85} to conductivity and the
density of states can also be obtained in this way.

\subsection{Renormalization group procedure}

In this section we work in two spatial dimensions and rewrite the model of
Eq.~(\ref{eq:ssigmamodel1}) in the following form: 
\begin{eqnarray}
&&F[Q] =\frac{\pi \nu }{4}\int d\mathbf{r}\;\mathrm{Str}\big[D(\nabla Q(%
\mathbf{r}))^{2}-4zEQ(\mathbf{r})\big]\qquad \\
&&+\frac{(\pi \nu )^{2}}{2}\int d\mathbf{r}d\mathbf{r^{\prime }}\left\langle 
\mathrm{Str}\big[\underline{{\hat{\phi}_{2}}}(\mathbf{r})Q(\mathbf{r})\big]\;%
\mathrm{Str}\big[\underline{{\hat{\phi}_{2}}}(\mathbf{r^{\prime }})Q(\mathbf{%
r^{\prime }})\big]\right\rangle _{\hat{\phi}_{2}}  \nonumber \\
&&-\frac{(\pi \nu )^{2}}{2}\int d\mathbf{r}d\mathbf{r^{\prime }}\left\langle 
\mathrm{Str}\big[\underline{{\hat{\phi}_{1}}}(\mathbf{r})Q(\mathbf{r})\big]\;%
\mathrm{Str}\big[\underline{{\hat{\phi}_{1}}}(\mathbf{r^{\prime }})Q(\mathbf{%
r^{\prime }})\big]\right\rangle _{\hat{\phi}_{1}}.  \nonumber
\label{renmodel}
\end{eqnarray}%
In order to reduce the number of terms we use the fields $\hat{\phi}%
_{i}=\phi _{i}+\varphi _{i}$ together with the symbolic notation $%
\left\langle \hat{\phi}_{i}\hat{\phi}_{i}\right\rangle =\left\langle \phi
_{i}\phi _{i}\right\rangle +\left\langle \varphi _{i}\varphi
_{i}\right\rangle $. The ``charge'' $z$ has been introduced as discussed
before. We start the renormalization group procedure separating slow and
fast modes 
\[
Q=UQ_{0}\bar{U},\qquad Q_{0}=U_{0}\Lambda \bar{U}_{0},\quad U\bar{U}=U_{0}%
\bar{U}_{0}=1. 
\]%
Fluctuations described by the supermatrix $U_{0}$ are fast while those
described by $U$ are slow. As a result of this separation we write the free
energy $F\left[ Q\right] $ in the form $F\left[ Q\right] =F^{(0)}+F_{int}$,
where 
\begin{eqnarray}
F^{\left( 0\right) } &=&\frac{\pi \nu }{4}\int d\mathbf{r}\;\mathrm{Str}\big[%
D(\nabla Q_{0})^{2}+2D\Phi \lbrack Q_{0},\nabla Q_{0}]  \nonumber \\
&&\qquad +D[Q_{0},\Phi ]^{2}-4zEUQ_{0}\bar{U}\big], \\
F_{int} &=&-\frac{(\pi \nu )^{2}}{2}\int d\mathbf{r}d\mathbf{r^{\prime }}%
\;\left( \left\langle \mathrm{Str}\big[\underline{{\hat{\phi}_{1\mathbf{r}}}}%
U_{\mathbf{r}}Q_{0\mathbf{r}}\bar{U}_{\mathbf{r}}\big]\right. \right. 
\nonumber \\
&\times &\left. \left. \mathrm{Str}\big[\underline{{\hat{\phi}_{1\mathbf{%
r^{\prime }}}}}U_{\mathbf{r^{\prime }}}Q_{0\mathbf{r^{\prime }}}\bar{U}_{%
\mathbf{r^{\prime }}}]\big]\right\rangle _{\hat{\phi}_{1}}-\left( \hat{\phi}%
_{1}\leftrightarrow \hat{\phi}_{2}\right) \right) ,\quad
\end{eqnarray}%
and $\mathbf{\Phi }=\bar{U}\mathbf{\nabla }U$.

Next, we parametrize fast modes $Q_{0}$ in the following way: 
\[
Q_{0}=\Lambda (1+iP)(1-iP)^{-1},\quad \{P,\Lambda \}=0. 
\]%
One must specify the precise momentum range related to the fast and slow
modes and additionally account for the frequency dependence. This can be
done in the following way:

\begin{enumerate}
\item Frequencies in the interval $\lambda \tau^{-1}<|\omega_n|<\tau^{-1}$, $%
0<\lambda<1$ and momenta in the shell $\lambda\tau^{-1}<Dk^2/z<\tau^{-1}$
are referred to as fast.

\item If slow variables $U$ have at least one Matsubara index corresponding
to the fast frequencies, they must be set equal to $\mathbbm{1}$.

\item If fast variables $P$ do not have at least one fast frequency or fast
momentum, they vanish.
\end{enumerate}

The separation of modes in the energy part requires some care. The energy
matrix must be split into a fast and a slow part: $E=E^{f}+E^{s}$. Then 
\begin{eqnarray}
\int d\mathbf{r}\;\mathrm{Str}\big[EU\Lambda P^{2}\bar{U}\big] &=&\int d%
\mathbf{r}\;\mathrm{Str}\big[E^{s}U\Lambda P^{2}\bar{U}\big]  \nonumber \\
&+&\int d\mathbf{r}\;\mathrm{Str}\big[E^{f}\Lambda P^{2}\big].
\end{eqnarray}%
Using the parametrization introduced above the free energy splits into
several parts. First, the original free energy is recovered with $Q$
replaced by $\tilde{Q}=U\Lambda \bar{U}$ and with the slow energy matrix $%
E^{s}$. In Finkel'steins scheme it is sufficient to keep terms of the second
order in $P$ for the remaining parts. As a result one finds%
\begin{equation}
F^{\left( 0\right) }=F_{0}+F_{1}+F_{2}+F_{E},  \label{e19}
\end{equation}%
\begin{eqnarray}
F_{0} &=&\pi \nu \int d\mathbf{r}\;\mathrm{Str}\big[D\nabla P\nabla
P+2zE^{f}\Lambda P^{2}\big], \\
F_{1} &=&2\pi \nu D\int d\mathbf{r}\;\mathrm{Str}\big[\Phi \lbrack P,\nabla
P]\big], \\
F_{2} &=&-2\pi \nu D\int d\mathbf{r}\;\mathrm{Str}\big[(\Phi \Lambda
)^{2}P^{2}+(\Lambda P\Phi )^{2}\big], \\
F_{E} &=&2\pi \nu z\int d\mathbf{r}\;\mathrm{Str}\big[E^{s}U\Lambda P^{2}%
\bar{U}\big].
\end{eqnarray}%
The interaction part $F_{int}=F_{int,1}+F_{int,2}$ can be written as follows: 
\begin{eqnarray}
&&F_{int,1}=2(\pi \nu )^{2}\int d\mathbf{r}d\mathbf{r^{\prime }}\;\left(
\left\langle \mathrm{Str}\big[\underline{{\hat{\phi}_{1\mathbf{r}}}}U_{%
\mathbf{r}}\Lambda P_{\mathbf{r}}\bar{U}_{\mathbf{r}}\big]\right. \right. 
\nonumber \\
&&\left. \left. \times \mathrm{Str}\big[\underline{{\hat{\phi}_{1\mathbf{%
r^{\prime }}}}}U_{\mathbf{r^{\prime }}}\Lambda P_{\mathbf{r^{\prime }}}%
\bar{U}_{\mathbf{r^{\prime }}}\big]\right\rangle _{\hat{\phi}_{1}}-\left( 
\hat{\phi}_{1}\leftrightarrow \hat{\phi}_{2}\right) \right), \\
&&F_{int,2}=2(\pi \nu )^{2}\int d\mathbf{r}d\mathbf{r^{\prime }}\;\left(
\left\langle \mathrm{Str}\big[\underline{{\hat{\phi}_{1\mathbf{r}}}}Q_{%
\mathbf{r}}\big]\right. \right.  \nonumber \\
&&\left. \left. \times \mathrm{Str}\big[\underline{{\hat{\phi}_{1\mathbf{%
r^{\prime }}}}}U_{\mathbf{r^{\prime }}}\Lambda P_{\mathbf{r^{\prime }}}^{2}%
\bar{U}_{\mathbf{r^{\prime }}}\big]\right\rangle _{\hat{\phi}_{1}}-\left( 
\hat{\phi}_{1}\leftrightarrow \hat{\phi}_{2}\right) \right).
\end{eqnarray}%
The fast modes can be integrated out in the Gaussian approximation and an
effective free energy $\tilde{F}$ is generated in this way 
\begin{eqnarray}
&&\tilde{F}[\tilde{Q}]=-\ln \left( \int DP\;\mbox{e}%
^{-F_{1}-F_{2}-F_{E}-F_{int}}\;\mbox{e}^{-F_{0}}\right) +F[\tilde{Q}]. 
\nonumber \\
&&
\end{eqnarray}%
The functional $\tilde{F}$ is the appropriate free energy for the slow
modes, taking into account the influence of the fast modes. It turns out
that this influence can effectively be expressed by a change of $D,z,\Gamma
_{1}$, and $\Gamma _{2}$ with the scale determined by $\lambda $ and this is
what demonstrates the renormalizability. From now on it will be clear from
the context whether we refer to slow $\tilde{Q}$ or fast $Q$, so that we can
simply use a common symbol $Q$.

\subsection{Diffusion coefficient}

\label{ssec:Dren} Here we calculate the corrections to the diffusive part 
\[
F_{D}=\frac{\pi \nu }{4}\int \mathrm{Str}\left[ D(\nabla Q)^{2}\right]. 
\]%
The energy part does not contribute and the renormalization of the diffusive
part can be written in the form 
\begin{eqnarray}
\delta {F_{D}} &=&\left\langle F_{int}\right\rangle -\left\langle
\!\left\langle F_{1}F_{int}\right\rangle \!\right\rangle -\left\langle
\!\left\langle F_{2}F_{int}\right\rangle \!\right\rangle +\frac{1}{2}%
\left\langle \!\left\langle F_{1}^{2}F_{int}\right\rangle \!\right\rangle . 
\nonumber  \label{eq:deltaFD} \\
&&
\end{eqnarray}%
By $\left\langle \!\left\langle \dots \right\rangle \!\right\rangle $ we
denote the connected part of the averages.

The term $F_{int}$ itself does not contain any gradients. It is therefore
not obvious, how $\left\langle F_{int}\right\rangle $ can contribute to the
renormalization of the diffusion coefficient. Therefore we discuss this
calculation in some detail.

According to the contraction rule (\ref{contract2}) we have $\left\langle
PP\right\rangle =0$. As a direct consequence of this property the average $%
\left\langle F_{int,2}\right\rangle $ vanishes. For the part $F_{int,1}$ we
should distinguish between several possible cases.

\begin{enumerate}
\item[(a)] The matrices $P$ can have two fast frequency indices. Then the $U$%
-modes completely vanish from the expression. It is quadratic in $P$ and
therefore contributes together with $F_{0}$ to the quadratic form in $P$.
This leads to the dressing of the diffusion modes in Finkel'stein's
approach. In our model these higher order interaction effects can only be
taken into account in special cases (as in Sec.~\ref{sec:perturbationtheory}),
while in general we have to restrict ourselves to the lowest order in the
interaction amplitudes for the renormalization.

\item[(b)] The matrices $P$ have two slow frequency indices, but the
momentum is fast. It turns out that integrating over $P$ the term is
quadratic in the slow fields $Q$ and contributes to the renormalization of
the interaction part (see Sec.~\ref{ssec:intren} below).

\item[(c)] The matrices $P$ have one fast and one slow frequency index. The
diffusion propagator, arising after calculating Gaussian integrals in $P$
can be expanded in slow momenta, which leads to a correction to the
diffusion coefficient. If it is expanded in small frequencies, it
contributes to the renormalization of $z$. The latter case becomes important
only if the diffusion modes are dressed and we refer to Ref.~\cite%
{Finkelstein90} for details. The former case will be discussed next.
\end{enumerate}

We consider the case (c). After integration over $P$-modes one obtains 
\begin{eqnarray}
\left\langle F_{int}\right\rangle &=&-2\pi \nu T^{2}\int d\mathbf{r}d\mathbf{%
r^{\prime }}\;\mathcal{D}(\mathbf{r}-\mathbf{r^{\prime }},k-l)  \nonumber \\
&&\times \mathrm{Str}\left[ \left( \left\langle \hat{\phi}_{1\mathbf{%
r^{\prime }}}^{-j_{2}}\hat{\phi}_{1\mathbf{r}}^{j_{1}}\right\rangle _{\hat{%
\phi}_{1}}-\left\langle \hat{\phi}_{2\mathbf{r^{\prime }}}^{-j_{2}}\hat{\phi}%
_{2\mathbf{r}}^{j_{1}}\right\rangle _{\hat{\phi}_{2}}\right) \right. 
\nonumber \\
&&\qquad \qquad \times \left. \left( \Delta ^{j_{1}}U_{\mathbf{r}}\right)
_{kl}^{\perp }\left( \bar{U}_{\mathbf{r^{\prime}}}\Delta ^{-j_{2}}\right)
_{lk}^{\perp }\right] ,\quad
\end{eqnarray}%
where $k$ is a fast frequency index while $l$ is slow. In Sec.~\ref%
{sec:smodel} it has been stressed that Eqs.~(\ref{unity1}) and (\ref{unity2})
are crucial for the model. It is in fact very important for the consistency
of the theory, that $\left\langle \hat{\phi}\hat{\phi}\right\rangle \propto %
\mathbbm{1}$. Integrating over $\hat{\phi}$ we find 
\begin{eqnarray}
&&\left\langle F_{int}\right\rangle =-2\pi T\;(\Gamma _{1}-2\Gamma
_{2})\;\int (d\mathbf{p})(d\mathbf{q})\;\mathcal{D}(\mathbf{p}+\mathbf{q}%
,k-l)  \nonumber \\
&&\times \mathrm{Str}\left[ U_{\mathbf{q},ml}^{\perp }\bar{U}_{-\mathbf{q}%
,lm}^{\perp }\Theta _{k,m}+U_{\mathbf{q},ml}^{\parallel }\bar{U}_{-\mathbf{q}%
,lm}^{\parallel }\Theta _{-k-1,m}\right].\no\\
\end{eqnarray}%
Here $\mathbf{p}$ is a fast momentum while $\mathbf{q}$ is slow. Keeping
only fast $k$ and $\mathbf{p}$ in the diffusion propagator the $U$
dependence would drop completely. Therefore one expands the diffusion
propagator up to second order in $\mathbf{q}$. Using the identity $\mathrm{%
Str}[\nabla U\nabla \bar{U}]=-\mathrm{Str}[\Phi ^{2}]$ and approximating the
frequency summation by an integral, we come to the following expression 
\begin{eqnarray}
&&\left\langle F_{int}\right\rangle =(\Gamma _{1}-2\Gamma _{2})\;\int d%
\mathbf{r}\;\mathrm{Str}\left[ \Phi ^{2}\right] \int (d\mathbf{p}) \nonumber \\
&&\times \int^{\prime }\!d\Omega \left( 2D^{2}\mathbf{p}%
^{2}\mathcal{D}^{3}(\mathbf{p},\Omega )-D\mathcal{D}^{2}(\mathbf{p},\Omega
)\right) .  \label{eq:FintforD}
\end{eqnarray}%
The momenta and frequencies in the region of the integration in Eq. (\ref%
{eq:FintforD}) are large and correspond to the fast variables. We use the
symbol $\int^{\prime }$ to indicate that integration is carried out over
positive frequencies only.

The remaining terms in Eq.~(\ref{eq:deltaFD}) can be considered in an
analogous way. The results are 
\begin{eqnarray}
&&-\left\langle\!\left\langle F_1F_{int}\right\rangle\!\right\rangle=
2(\Gamma_1-2\Gamma_2) \int d\mathbf{r}\;\mathrm{Str}\big[\Phi^{\parallel 2}%
\big]\;\int (d\mathbf{p})  \nonumber \\
&&\quad\times \int^{\prime}\!\! d\Omega \left(D%
\mathcal{D}^2(\mathbf{p},\Omega)-2D^2\mathbf{p}^2\mathcal{D}^3(\mathbf{p}%
,\Omega)\right),\quad\;  \label{eq:F1Fint} \\
&&-\left\langle\!\left\langle F_2F_{int}\right\rangle\!\right\rangle =-
(\Gamma_1-2\Gamma_2) \int d\mathbf{r}\;\mathrm{Str}\left[\Phi^{\parallel
2}-\Phi^{\perp 2}\right]  \nonumber \\
&&\qquad\qquad\times \int(d\mathbf{p})\!\! \int^{\prime}\!\! d\Omega\; D 
\mathcal{D}^2(\mathbf{p},\Omega),  \label{eq:F2Fint} \\
&&\frac{1}{2}\;\left\langle\!\left\langle
F_1^2F_{int}\right\rangle\!\right\rangle= 2(\Gamma_1-2\Gamma_2) \;\int d%
\mathbf{r}\; \mathrm{Str}\big[\Phi^{\parallel 2}\big]  \nonumber \\
&&\qquad\qquad\times\int (d\mathbf{p})\!\! \int^{\prime}\!\!d\Omega \;D^2%
\mathbf{p}^2\;\mathcal{D}^3(\mathbf{p},\Omega).  \label{eq:F12Fint}
\end{eqnarray}
Following Finkel'stein, terms that are parametrically small in $t D\mathbf{q}%
^2/(z\lambda\tau^{-1})\ll 1$, where $\mathbf{q}$ is a typical slow momentum,
have not been taken into account. Such terms can arise from expressions in
which all frequency indices are fixed to be slow by $\Phi$-modes\cite%
{Finkelstein90}. A comparison of the combined contribution of Eqs.~(\ref%
{eq:FintforD}--\ref{eq:F12Fint}) to the original diffusive part $F_D$ gives 
\begin{eqnarray}
\delta D&=&-\frac{1}{(2\pi)^2\nu}J_2,  \label{eq:deltaD} \\
J_2&=&8\pi (\Gamma_1-2\Gamma_2) \int(d\mathbf{p})\!\!\int^{\prime}\!\!d%
\Omega\;\frac{ D^2\mathbf{p}^2}{(D\mathbf{p}^2+z\Omega)^3}.\qquad
\end{eqnarray}

\subsection{Effective charge $z$}

\label{ssec:zren} As we have mentioned, in the $\sigma $-model for
interacting systems a new effective charge $z$ appears as a coefficient in
the original $\sigma$-model, 
\[
F_{z}=-\pi \nu z\int \mathrm{Str}\big[EQ\big]. 
\]%
Expanding this term in $P$ we write the relevant term of the second order in 
$P$ as 
\[
F_{E}=2\pi \nu z\int \mathrm{Str}\big[EU\Lambda P^{2}\bar{U}]. 
\]%
It is understood that the energy matrix $E$ contains only slow indices here.
In our case the correction to $z$ originates only from one term 
\[
\delta F_{z}=-\left\langle \!\left\langle F_{E}F_{int}\right\rangle
\!\right\rangle. 
\]%
Working with dressed interaction lines as in Ref.~\cite{Finkelstein90}, one
more contribution from $\left\langle F_{int}\right\rangle $ should be taken
into account. The final result, however, is the same in both cases and we
find 
\begin{eqnarray}
&&-\left\langle \!\left\langle F_{E}F_{int}\right\rangle \!\right\rangle
\sim z(\Gamma _{1}-2\Gamma _{2})  \nonumber \\
&&\times \;\int d\mathbf{r}\;\mathrm{Str}\big[EQ\big]\int (d\mathbf{p}%
)\!\!\int^{\prime }\!\!d\Omega \;\mathcal{D}^{2}(\mathbf{p},\Omega ).
\end{eqnarray}%
Equations~(\ref{unity1}) and (\ref{unity2}) were again crucial to arrive at this
result. The relation $\partial _{\Omega }\mathcal{D}(\mathbf{p},\Omega )=-z%
\mathcal{D}^{2}(\mathbf{p},\Omega )$ can be used now to obtain 
\begin{equation}
\delta z=\frac{1}{\pi \nu }\int (d\mathbf{p})\;(\Gamma _{1}-2\Gamma _{2})\;%
\left[ \mathcal{D}(\mathbf{p},\Omega )\right] _{\Omega =\lambda \tau
^{-1}}^{\Omega =\tau ^{-1}}.  \label{eq:deltaz}
\end{equation}%
Equation (\ref{eq:deltaz}) gives the logarithmic correction to the effective
charge that will be used for writing renormalization group equations.

\subsection{Interaction part}

\label{ssec:intren} The leading order corrections in $\Gamma _{1}$, $\Gamma
_{2}$ contributing to the renormalization of the interaction part come from
the average $\left\langle F_{int}\right\rangle $. We will see that it is
very important here to work with \emph{both} fields $\phi $ and $\varphi 
$. As pointed out above in Sec.~\ref{ssec:Dren}, for this term \emph{both} frequencies of $P$ should be slow,
 while the momentum is to be fast.
After integration over $P$ we obtain
\begin{eqnarray}
&&\left\langle F_{int}\right\rangle =\frac{\pi \nu }{2}\int (d\mathbf{p})\;%
\mathcal{D}(\mathbf{p},0)  \nonumber \\
&&\times \int d\mathbf{r}d\mathbf{r^{\prime }}\;\left\langle \mathrm{Str}%
\big[Q_{\mathbf{r}}\underline{\hat{\phi}_{1\mathbf{r}}}Q_{\mathbf{r}}%
\underline{\hat{\phi}_{1\mathbf{r^{\prime }}}}-Q_{\mathbf{r}}\underline{\hat{%
\phi}_{2\mathbf{r}}}Q_{\mathbf{r}}\underline{\hat{\phi}_{2\mathbf{r^{\prime }%
}}}\big]\right\rangle _{\hat{\phi}_{1},\hat{\phi}_{2}}  \nonumber \\
&=&\frac{1}{4}\int (d\mathbf{p})\;\mathcal{D}(\mathbf{p},0)\int d\mathbf{r}%
\;\left( \Gamma _{1}\;Q\gamma _{2}Q-\Gamma _{2}Q\gamma _{1}Q\right).
\end{eqnarray}%
We write $\mathcal{D}(\mathbf{p},0)$ here to symbolize that only the
momentum should be integrated, since frequencies of $P$-modes are slow. As a
result, the corrections to $\Gamma _{1}$ and $\Gamma _{2}$ are 
\begin{eqnarray}
\delta \Gamma _{1} &=&\frac{1}{\pi \nu }\;\Gamma _{2}\;\int (d\mathbf{p})\;%
\mathcal{D}(\mathbf{p},0),  \label{eq:deltaGamma1} \\
\delta \Gamma _{2} &=&\frac{1}{\pi \nu }\;\Gamma _{1}\int (d\mathbf{p})\;%
\mathcal{D}(\mathbf{p},0).  \label{eq:deltaGamma2}
\end{eqnarray}%
This agrees with the corresponding result obtained from the replica model %
\cite{Finkelstein90}, since the dressing of the diffusion mode is not
effective here. We would like to
stress that the correction to $\Gamma _{1}$ is proportional to $\Gamma _{2}$
and vice versa. From the technical point of view it is of particular
importance here that the terms $Q\gamma _{1}Q$ and $Q\gamma _{2}Q$ in the
supersymmetric model of Eq.~(\ref{eq:ssigmamodel1}) include both the single
supertrace and the product of two supertraces. Only in this way the correct
structure can be reproduced in the calculation of $\left\langle
F_{int}\right\rangle $.

\subsection{RG equations}

From Eqs.~(\ref{eq:deltaD}), (\ref{eq:deltaz}), (\ref{eq:deltaGamma1}), and (%
\ref{eq:deltaGamma2}) appropriate renormalization group equations may be
derived. They take the form 
\begin{eqnarray}
\frac{1}{t}\frac{dt}{d\ln \lambda ^{-1}} &=&\frac{t}{z}(\Gamma _{1}-2\Gamma
_{2}),  \label{grenorm} \\
\frac{1}{t}\frac{dz}{d\ln \lambda ^{-1}} &=&-(\Gamma _{1}-2\Gamma _{2}), 
\nonumber \\
\frac{1}{t}\frac{d\Gamma _{1}}{d\ln \lambda ^{-1}} &=&\Gamma _{2},  \nonumber
\\
\frac{1}{t}\frac{d\Gamma _{2}}{d\ln \lambda ^{-1}} &=&\Gamma _{1}.  \nonumber
\end{eqnarray}%
These equations are valid in the first order in $t=((2\pi )^{2}\nu D)^{-1}$
and in the leading order in the interaction amplitudes for $\Gamma _{i}\ll z$%
. In the first order in $\Gamma _{1,2}$ they coincide with the corresponding
limit of the renormalization group equations obtained in Refs.~\cite%
{Castellani83} and \cite{Chamon99} for short range interactions, where terms of all
orders in the interaction amplitudes were included. As our supersymmetric
model, Eqs. (\ref{e4}) and (\ref{e6}), can serve as a good approximation for the
initial electron model only for a weak interaction, taking into account the
terms of higher orders in interaction would lead to an overestimation.
Therefore, we conclude that in the region of applicability the
supersymmetric $\sigma $-model gives the correct renormalization group
equations.

From these equations one can see that the combination $z_{1}=z-2\Gamma
_{1}+\Gamma _{2}$ is invariant under renormalization. 
\begin{equation}
\frac{dz_{1}}{d\ln \lambda ^{-1}}=\frac{d}{d\ln \lambda ^{-1}}(z-2\Gamma
_{1}+\Gamma _{2})=0  \label{e23}
\end{equation}%
This invariance is not incidental but protected by a Ward identity and
therefore its validity is not restricted to the approximations studied here.
For more details we refer to Refs.~\cite%
{Finkelstein83,Castellani83,DiCastro03}, and \cite{Chamon99}. Its importance for the
consistency of the entire scheme will become obvious in the next section,
when we discuss once more the density-density correlation function.

\subsection{Conductivity}

\label{ssec:conductivity} During the process of the renormalization the
meaning of the coefficient $D$ appearing in the free energy and its relation
to conductivity remains not very well understood. In order to clarify its
meaning one can study the density-density correlation function \cite%
{Finkelstein83,Castellani83,Finkelstein90,DiCastro03}. In order to calculate
the density-density correlation functions one can add to the free energy
functional a source term of the form $F_{\vartheta }=i\pi \nu \int d\mathbf{r%
}\;\mathrm{Str}[\underline{\vartheta }Q]$, Eq. (\ref{e21}). Then corrections
to this term can be calculated. In our approximation one should consider the
average 
\[
-\left\langle \!\left\langle F_{\vartheta }F_{int,1}\right\rangle
\!\right\rangle . 
\]%
This expression covers vertex corrections as well as corrections to the
single particle Green function. These contributions exactly cancel, so that $F_{\vartheta }$, Eq. (\ref{e21}%
) is unchanged. [This is also true in all orders in $\Gamma _{i}$ \cite%
{Finkelstein90}]. We would like to emphasize that one may not use $%
F_{\varphi },$ Eq. (\ref{e22}), in this procedure because this would lead to
spurious contributions.

Now the dynamic part of the polarization can be calculated from the
renormalized $\sigma $-model as in Sec.~\ref{sec:perturbationtheory}. It is
generally expected that the thermodynamic density of states $\partial _{\mu
}n$ does not develop logarithmic singularities\cite%
{Finkelstein83,Castellani83,Finkelstein90,Belitz94}. Then we come to the
result 
\[
\hat{\Pi}(\mathbf{q},\Omega _{k})=-2\nu \;\frac{|\Omega _{k}|}{%
D\bfq^{2}+z_{1}|\Omega _{k}|}, 
\]%
where $D$ is the renormalized diffusion coefficient. The coefficient $z_{1}$
is a constant under renormalization as we have seen above, Eq. (\ref{e23}),
and using Eq. (\ref{pi}) we find again 
\[
\Pi (\mathbf{q},\Omega _{k})=\partial _{\mu }n\frac{D_{e}\bfq^2}{%
D_{e}\bfq^{2}+|\Omega _{k}|}, 
\]%
$D_{e}=(1+F_{0}^{s})D$. With the help of the equation of continuity (\ref%
{eq:continuity}) one recovers the Einstein relation 
\[
\sigma =e^{2}\partial _{\mu }nD_{e}. 
\]%
So, this relation remains valid and one can determine the conductivity from
the renormalized diffusion coefficient. As an example, one can make contact
to the Altshuler-Aronov corrections to the conductivity for the case of \
short range interactions. To this end one should replace $z$ by its bare
value $z=1$ as well as $\Gamma _{i}\rightarrow V_{i}$ in Eq.~(\ref{grenorm}%
), where $V_{1}=\nu V_{0}(0)$ and $V_{2}=\nu \overline{V_{0}(\mathbf{p}-%
\mathbf{p^{\prime }})}$ (the bar means averaging over the Fermi surface). As
a result one comes to the well-known formula, Eq. (\ref{e25})\cite%
{Altshuler85}.

\subsection{Density of states}

\label{sec:dos} The main purpose of this section is to obtain an expression
for interaction corrections to the density of states from the supersymmetric 
$\sigma $-model derived. The single particle density of states (per spin degree of freedom) can be
written as 
\[
\nu (\varepsilon )=-\frac{1}{\pi }\int (d\mathbf{p})\Im {G^{R}_{\sigma\sigma}(\mathbf{p}%
,\varepsilon )}. 
\]%
No summation over spin index $\sigma$ is implied.
In order to calculate the density of states in the Matsubara formalism it is
convenient to first define the function 
\[
\tilde{\nu}(\omega _{n})=-\frac{1}{\pi }\int d\tau \;\mathcal{G}_{\sigma\sigma}(\mathbf{r},%
\mathbf{r},\tau ,0)\;\mbox{e}^{i\omega _{n}\tau }. 
\]%
where $\mathcal{G}$ denotes the Matsubara Green function of the system. It
is sufficient to know $\tilde{\nu}(\omega _{n})$ for positive frequencies.
The density of states $\nu $ can be obtained from $\tilde{\nu}$ by an
analytic continuation, 
\[
\nu (\epsilon )=\Im \left( \tilde{\nu}\left. (\omega _{m})\right| _{\omega
_{m}\rightarrow -i\epsilon +\delta }\right) .
\]%
In our formalism we obtain (compare Eq.~\ref{e8})
\be
\tilde{\nu}(\omega _{m})=-i\nu \;\left\langle\mathrm{Str}\big[k_{+}Q_{mm}^{\sigma\sigma}(\mathbf{r})]\right\rangle,
\ee
assuming that the saddle point approximation has been employed.
Again, we separate fast and slow modes in $Q$ and expand $Q_0$ up to second order in $P$. 
We also expand $\exp(-F)$ up to first order in $F_{int}$. We should consider the correction
\be
\delta\tilde{\nu}=i\nu\left\langle\!\left\langle \;\mathrm{Str}\big[k_{+}Q_{mm}^{\sigma\sigma}(\mathbf{r})]  
F_{int}\right\rangle\!\right\rangle.
\ee 
The calculation is very similar to that of $z$ and one obtains the correction to the
density of states in the form 
\begin{equation}
\delta\nu=-\nu \frac{t}{z} (\Gamma_1-2\Gamma_2)\ln\lambda^{-1}.
  \label{e27}
\end{equation}

Equation (\ref{e27}) allows to write a renormalization group
equation for the density of states 
\begin{equation}
\frac{1}{\nu}\frac{\partial \nu}{\partial \ln\lambda^{-1}}=-\frac{t}{z} (\Gamma_1-2\Gamma_2).
  \label{e28}
\end{equation}
The validity is, of course, restricted to small effective interaction amplitudes $\Gamma_i/z\ll 1$.
As in Sec.~\ref{ssec:conductivity} one can recover the Altshuler-Aronov
correction for the short range case that can be written in the standard form 
\begin{equation}
\delta\nu/\nu= -t(V_1-2V_2)\ln (T\tau).
\label{e26}
\end{equation}
The same formulas Eqs. (\ref{e27})--(\ref{e26}) for the density of states would be obtained with Replica or Keldysh models in the limit of weak interactions.

\subsection{Discussion of the renormalization group}

The renormalization analysis carried out in this section allows us to come
to the conclusion that the supersymmetric $\sigma $-model is renormalizable
at least in the first order in the interaction amplitudes and inverse conductance $%
t=((2\pi )^{2}\nu D)^{-1}.$ The structure of the interaction term, Eqs. (\ref%
{eq:gamma1}) and (\ref{eq:gamma2}) is crucial for the renormalizability and
therefore the renormalization group treatment can serve as a very good check
of the $\sigma $-model. The renormalization group equations derived above
(Eq.~(\ref{grenorm})) agree with those given in Refs.~\cite%
{Castellani83,Finkelstein90,Chamon99}, and \cite{DiCastro03} in the leading order in $%
\Gamma _{i}$ and for $\Gamma _{i}\ll z$.

We do not present here the solution of these equations referring the reader
to Refs.\cite{Castellani83,Finkelstein90,Chamon99}, and\cite{DiCastro03} where
solutions of more general equations have been discussed for arbitrary values
of the amplitudes $\Gamma _{i}$. Unfortunately, even these more general
equations do not allow for definite conclusions in two dimensions, since
under renormalization the quotient $\Gamma _{2}/z$ diverges. This leads to a
divergence of the spin susceptibility, which has been interpreted as some
sort of ferromagnetic instability \cite{Finkelstein90}. However, it should
be stressed once again that the renormalization group calculations were
performed here only as a check of the new $\sigma $-model. For the more
complete study of the scaling behavior in $2D$ replica or Keldysh models are
more convenient. The remarkable point is that the nontrivial set of
Eqs. (\ref{grenorm}) has been obtained from a model based entirely on
the supersymmetry method.

\section{Discussion}

In this paper we constructed a supersymmetric $\sigma $-model for disordered
fermion systems with interaction. Instead of trying to combine the
supersymmetry technique with an electron-electron interaction for the
initial electron model we proposed an artificial fully supersymmetric model
with interaction. The derivation of the $\sigma $-model from the latter
model is straightforward and valid for an arbitrary interaction.
Unfortunately, the artificial supersymmetric model is equivalent to the
electron one only in the limit of a weak short range interaction. Therefore
the $\sigma $-model we obtained is less general than the replica or Keldysh $%
\sigma $-models. So, as concerns the perturbation theory in the diffusion
modes or the renormalization group calculations, it is better to perform them
with the replica or Keldysh $\sigma $-models and our goal was not to
construct a more convenient tool for such calculations. Although we have
demonstrated how to carry out perturbative and RG computations in the
framework of our $\sigma $-model, the purpose of these calculations was
merely to check the $\sigma $-model. The check is successful and our hope is
that using the supersymmetric $\sigma $-model we derived one can perform
nonperturbative calculations.

It is well known that for noninteracting systems the supersymmetry
technique has real advantages with respect to the other methods \cite%
{Efetov97} and allows one to consider, e.g., localization in wires, level
statistics in quantum dots, etc. So, our hope is that the supersymmetric $%
\sigma $-model with interaction derived in the present paper can help us to
get results in this direction. The supermatrices $Q$ derived here should
have the correct symmetry (a mixture between compact and noncompact
sectors) and there is no reason to think that, when properly treated, the
model would lead to wrong results.

Nevertheless, using the present $\sigma $-model is definitely more difficult
than the conventional $\sigma $-model for noninteracting particles. The
problem is that the supermatrix $Q$ is now not a $8\times 8$ or $4\times 4$
supermatrix as it was for the noninteracting systems. In contrast, the size
of the supermatrix $Q$ is now $4M\times 4M$ (for the unitary ensemble),
where $M$ is the number of the Matsubara frequencies involved in the
calculation (actually, it is infinitely large). Therefore, we should be able
to calculate integrals over the supermatrices of an arbitrary large size. Of
course, the same problem exists for the replica and Keldysh $\sigma $-model.
In the replica $\sigma $-model one should, in addition, calculate for an
arbitrary number $N$ of the replicas and set at the end $N=0$. In the
Keldysh approach the matrix $Q$ depends on two energies and one should
discretize the energy to get a reasonable result.

The advantage of the supersymmetry approach is that the supersymmetric $%
\sigma $-model is still simpler and well defined.

In principle, the task of making explicit calculations with $4M\times 4M$
supermatrices does not seem hopeless. For calculations in the $0D$ situation
one can try to use the Itzykson-Zuber integral\cite{Itzykson80,Harish57} that
has been generalized for the supersymmetric case by Guhr et al.\cite{Guhr96}. An
integral with the zero-dimensional supermatrix free energy functional taken
in the form of Eq. (\ref{eq:fintwithphi})
has the proper form and one may hope to proceed in this way. Another problem
is localization in wires. With the supermatrix $\sigma $-model one can write
using the transfer matrix technique an effective ``Schr\"{o}dinger
equation'' and then try to solve it. These are the most evident examples.
Probably, one can try other situations, although the proper calculations can
be difficult.

Of course, at the moment we cannot guarantee the complete success in this
direction. At the same time, it is difficult to imagine that either replica
or Keldysh $\sigma $-models can be more helpful. Although there is a certain
progress in reproducing some known results for noninteracting systems using
these methods \cite{Kamenev99a,Yurkevich99,Altland00,Kanzieper02}, the calculations are
quite difficult and the results are not complete. We do not see how these
schemes can be extended to interacting systems and believe that the
supermatrix $\sigma $-model developed here is a better opportunity to attack
the nonperturbative problems.


\begin{acknowledgments}
The authors gratefully acknowledge the financial support of SFB TR 12 and GRK 384. The authors
would like to thank I. Aleiner, I. Beloborodov, A. Lopatin, and K. Takahashi for useful discussions at the initial stages of the work. 
\end{acknowledgments}

\appendix* 

\section{Parametrization of $\protect\phi_1$, $\protect\phi_2$ and $\protect%
\varphi_1$, $\protect\varphi_2$}

\label{appsec:parameterization} Here the definition of the fields $%
\phi_i,\varphi_i$ is given. In the following we write supermatrices with $%
2\times 2$ block structure describing spin space. $a^{ij},b^{ij}$ are
commuting while $\sigma^{ij}$ are anticommuting. Variables $a,b,\sigma$ for
different $\phi_i,\varphi_i$ are not related to one another. $a(\mathbf{r}%
,\tau),b(\mathbf{r},\tau)$ and $a^{ii}(\mathbf{r},\tau),b^{ii}(\mathbf{r}%
,\tau)$ are real. We assume $\Gamma_i$ to be short range and usually set $%
\Gamma_i(\mathbf{r}-\mathbf{r^{\prime}})=\Gamma_i \delta(\mathbf{r}-\mathbf{%
r^{\prime}})$. $\mathcal{N}$ are normalization constants for the Gaussian
integrals. Integration is over all independent components of the
corresponding matrices, 
\begin{widetext}
\be
\varphi_1(\bfr,n)&=&\left(\ba{cc} \ba{cc} a(\bfr,n)&0\\0&a(\bfr,n)\ea & 0\\0&\ba{cc} a(\bfr,n)&0\\0&a(\bfr,n)\ea\ea\right),\\
&&\no\\
\varphi_2(\bfr,n)&=&\left(\ba{cc} \ba{cc} a^{11}(\bfr,n)&a^{12}(\bfr,n) \\a^{12\, *}(\bfr,-n)&a^{22}(\bfr,n)\ea & 0\\0&\ba{cc} a^{11}(\bfr,n)&a^{12}(\bfr,n) \\a^{12\,* }(\bfr,-n)&a^{22}(\bfr,n)\ea   \ea\right),\\
&&\no\\
\phi_1(\bfr,n)&=&\left(\ba{cc}\ba{cc} a(\bfr,n)&0\\0&a(\bfr,n)\ea &\ba{cc} \sigma^*(\bfr,-n)&0\\0&\sigma^*(\bfr,-n)\ea  \\\ba{cc} \sigma(\bfr,n)&0\\0&\sigma(\bfr,n)\ea &\ba{cc} ib(\bfr,n)&0\\0&ib(\bfr,n)\ea\ea\right),\\
&&\no\\
\phi_2(\bfr,n)&=&\left(\ba{cc}\ba{cc} a^{11}(\bfr,n)&a^{12}(\bfr,n)\\a^{12\,*}(\bfr,-n)&a^{22}(\bfr,n)\ea &\ba{cc} \sigma^{11\,*}(\bfr,-n)&\sigma^{12\,*}(\bfr,-n)\\\sigma^{21\,*}(\bfr,-n)&\sigma^{22\,*}(\bfr,-n)\ea  \\\ba{cc} \sigma^{11}(\bfr,n)&\sigma^{21}(\bfr,n)\\\sigma^{12}(\bfr,n)&\sigma^{22}(\bfr,n)\ea &\ba{cc} ib^{11}(\bfr,n)&ib^{12}(\bfr,n)\\ib^{12\,*}(\bfr,-n)&ib^{22}(\bfr,n)\ea\ea\right),
\ee
and
\be
\left\langle\dots\right\rangle_{\varphi_1}&=&\mathcal{N}\int D\varphi_1\;(\dots)\;\mbox{e}^{-\frac{T}{8}\int d\bfr d\bfrp \;\tr\left[\varphi_1(\bfr,-n)\nu\Gamma^{-1}_1(\bfr-\bfrp)\varphi_1(\bfrp,n)\right]},\\
\left\langle\dots\right\rangle_{\varphi_2}&=&\mathcal{N}\int D\varphi_2\;(\dots)\;\mbox{e}^{-\frac{T}{4}\int d\bfr d\bfrp\; \tr\left[\varphi_2(\bfr,-n)\nu\Gamma_2^{-1}(\bfr-\bfrp)\varphi_2(\bfrp,n)\right]},\\
\left\langle\dots\right\rangle_{\phi_1}&=&\mathcal{N}\int D\phi_1\;(\dots)\;\mbox{e}^{-\frac{T}{4}\int d\bfr d\bfrp\;\Str\left[ \phi_1(\bfr,-n)\nu\Gamma_1^{-1}(\bfr-\bfrp)\phi_1(\bfrp,n)\right]},\\
\left\langle\dots\right\rangle_{\phi_2}&=&\mathcal{N}\int D\phi_2\;(\dots)\;\mbox{e}^{-\frac{T}{2}\int d\bfr d\bfrp\; \Str\left[ \phi_2(\bfr,-n)\nu\Gamma_2^{-1}(\bfr-\bfrp)\phi_2(\bfrp,n)\right]}.
\ee

\end{widetext}


\begin{thebibliography}{28}
\expandafter\ifx\csname natexlab\endcsname\relax\def\natexlab#1{#1}\fi
\expandafter\ifx\csname bibnamefont\endcsname\relax
  \def\bibnamefont#1{#1}\fi
\expandafter\ifx\csname bibfnamefont\endcsname\relax
  \def\bibfnamefont#1{#1}\fi
\expandafter\ifx\csname citenamefont\endcsname\relax
  \def\citenamefont#1{#1}\fi
\expandafter\ifx\csname url\endcsname\relax
  \def\url#1{\texttt{#1}}\fi
\expandafter\ifx\csname urlprefix\endcsname\relax\def\urlprefix{URL }\fi
\providecommand{\bibinfo}[2]{#2}
\providecommand{\eprint}[2][]{\url{#2}}

\bibitem[{\citenamefont{Wegner}(1979)}]{Wegner79}
\bibinfo{author}{\bibfnamefont{F.}~\bibnamefont{Wegner}}, \bibinfo{journal}{Z.
  Phys. B} \textbf{\bibinfo{volume}{35}}, \bibinfo{pages}{207}
  (\bibinfo{year}{1979}).

\bibitem[{\citenamefont{Sch\"afer and Wegner}(1980)}]{schaefer80}
\bibinfo{author}{\bibfnamefont{L.}~\bibnamefont{Sch\"afer}} \bibnamefont{and}
  \bibinfo{author}{\bibfnamefont{F.}~\bibnamefont{Wegner}},
  \bibinfo{journal}{Z. Phys. B:Condens. Matter} \textbf{\bibinfo{volume}{38}},
  \bibinfo{pages}{113} (\bibinfo{year}{1980}).


\bibitem[{\citenamefont{Efetov et~al.}(1980)\citenamefont{Efetov, Larkin, and
  Khmelnitskii}}]{Efetov80}
\bibinfo{author}{\bibfnamefont{K.~B.} \bibnamefont{Efetov}},
  \bibinfo{author}{\bibfnamefont{A.~I.} \bibnamefont{Larkin}},
  \bibnamefont{and} \bibinfo{author}{\bibfnamefont{D.~E.}
  \bibnamefont{Khmelnitskii}}, \bibinfo{journal}{Zh. Eksp. Teor. Fiz.}
  \textbf{\bibinfo{volume}{79}}, \bibinfo{pages}{1120} (\bibinfo{year}{1980}) [\bibinfo{journal}{Sov. Phys. JETP} \textbf{\bibinfo{volume}{52}}, \bibinfo{pages}{568} (\bibinfo{year}{1980})].


\bibitem[{\citenamefont{Efetov}(1982)}]{Efetov82}
\bibinfo{author}{\bibfnamefont{K.~B.} \bibnamefont{Efetov}},
  \bibinfo{journal}{Zh. Eksp. Teor. Fiz.} \textbf{\bibinfo{volume}{82}},
  \bibinfo{pages}{872} (\bibinfo{year}{1982}) [\bibinfo{journal}{Sov. Phys. JETP} \textbf{\bibinfo{volume}{55}}, \bibinfo{pages}{514} (\bibinfo{year}{1982})].

\bibitem[{\citenamefont{Efetov}(1997)}]{Efetov97}
\bibinfo{author}{\bibfnamefont{K.~B.} \bibnamefont{Efetov}},
  \emph{\bibinfo{title}{Supersymmetry in Disorder and Chaos}}
  (\bibinfo{publisher}{Cambridge University Press}, \bibinfo{address}{Cambridge},
  \bibinfo{year}{1997}).


\bibitem[{\citenamefont{Finkel'stein}(1983)}]{Finkelstein83}
\bibinfo{author}{\bibfnamefont{A.}~\bibnamefont{Finkel'stein}},
  \bibinfo{journal}{Zh. Eksp. Teor. Fiz.} \textbf{\bibinfo{volume}{84}},
  \bibinfo{pages}{168} (\bibinfo{year}{1983}) [\bibinfo{journal}{Sov. Phys. JETP} \textbf{\bibinfo{volume}{57}},
  \bibinfo{pages}{97} (\bibinfo{year}{1983})].


\bibitem[{\citenamefont{Keldysh}(1964)}]{Keldysh64}
\bibinfo{author}{\bibfnamefont{L.}~\bibnamefont{Keldysh}},
  \bibinfo{journal}{Zh. Eksp. Teor. Fiz.} \textbf{\bibinfo{volume}{47}},
  \bibinfo{pages}{1515} (\bibinfo{year}{1964}) [\bibinfo{journal}{Sov. Phys. JETP} \textbf{\bibinfo{volume}{20}}, \bibinfo{pages}{1018} (\bibinfo{year}{1965})].


\bibitem[{\citenamefont{Aronov and Ioselevich}(1985)}]{Aronov85}
\bibinfo{author}{\bibfnamefont{A.~G.} \bibnamefont{Aronov}} \bibnamefont{and}
  \bibinfo{author}{\bibfnamefont{A.~S.} \bibnamefont{Ioselevich}},
  \bibinfo{journal}{Pisma Zh. Eksp. Teor. Fiz.} \textbf{\bibinfo{volume}{41}},
  \bibinfo{pages}{71} (\bibinfo{year}{1985}) [\bibinfo{journal}{JETP Lett.} \textbf{\bibinfo{volume}{41}}, \bibinfo{pages}{84} (\bibinfo{year}{1985})].

\bibitem[{\citenamefont{Horbach and Sch\"on}(51)}]{Horbach93}
\bibinfo{author}{\bibfnamefont{M.}~\bibnamefont{Horbach}} \bibnamefont{and}
  \bibinfo{author}{\bibfnamefont{G.}~\bibnamefont{Sch\"on}},
  \bibinfo{journal}{Ann. Phys. (N.Y.)} \textbf{\bibinfo{volume}{2}},
  \bibinfo{pages}{51} (\bibinfo{year}{1993}).

\bibitem[{\citenamefont{Kamenev and Andreev}(1999)}]{Kamenev99}
\bibinfo{author}{\bibfnamefont{A.}~\bibnamefont{Kamenev}} \bibnamefont{and}
  \bibinfo{author}{\bibfnamefont{A.}~\bibnamefont{Andreev}},
  \bibinfo{journal}{Phys. Rev. B} \textbf{\bibinfo{volume}{60}},
  \bibinfo{pages}{2218} (\bibinfo{year}{1999}).

\bibitem[{\citenamefont{Chamon et~al.}(1999)\citenamefont{Chamon, Ludwig, and
  Nayak}}]{Chamon99}
\bibinfo{author}{\bibfnamefont{C.}~\bibnamefont{Chamon}},
  \bibinfo{author}{\bibfnamefont{A.}~\bibnamefont{Ludwig}}, \bibnamefont{and}
  \bibinfo{author}{\bibfnamefont{C.}~\bibnamefont{Nayak}},
  \bibinfo{journal}{Phys. Rev. B} \textbf{\bibinfo{volume}{60}},
  \bibinfo{pages}{2239} (\bibinfo{year}{1999}).

\bibitem[{\citenamefont{Mehta}(1991)}]{Mehta91}
\bibinfo{author}{\bibfnamefont{M.~L.} \bibnamefont{Mehta}},
  \emph{\bibinfo{title}{Random Matrices}} (\bibinfo{publisher}{Academic},
  \bibinfo{address}{San Diego}, \bibinfo{year}{1991}).

\bibitem[{\citenamefont{Verbaarschot et~al.}(1985)\citenamefont{Verbaarschot,
  Weidenm\"uller, and Zirnbauer}}]{Verbaarschot85a}
\bibinfo{author}{\bibfnamefont{J.~J.~M.} \bibnamefont{Verbaarschot}},
  \bibinfo{author}{\bibfnamefont{H.~A.} \bibnamefont{Weidenm\"uller}},
  \bibnamefont{and} \bibinfo{author}{\bibfnamefont{M.~R.}
  \bibnamefont{Zirnbauer}}, \bibinfo{journal}{Phys. Rep.}
  \textbf{\bibinfo{volume}{129}}, \bibinfo{pages}{367} (\bibinfo{year}{1985}).

\bibitem[{\citenamefont{Verbaarschot and Zirnbauer}(1985)}]{Verbaarschot85}
\bibinfo{author}{\bibfnamefont{J.~J.~M.} \bibnamefont{Verbaarschot}}
  \bibnamefont{and} \bibinfo{author}{\bibfnamefont{M.~R.}
  \bibnamefont{Zirnbauer}}, \bibinfo{journal}{J. Phys. A}
  \textbf{\bibinfo{volume}{18}}, \bibinfo{pages}{1093} (\bibinfo{year}{1985}).

\bibitem[{\citenamefont{Kamenev and Mezard}(1999)}]{Kamenev99a}
\bibinfo{author}{\bibfnamefont{A.}~\bibnamefont{Kamenev}} \bibnamefont{and}
  \bibinfo{author}{\bibfnamefont{M.}~\bibnamefont{Mezard}},
  \bibinfo{journal}{J. Phys. A} \textbf{\bibinfo{volume}{32}},
  \bibinfo{pages}{4373} (\bibinfo{year}{1999}).

\bibitem[{\citenamefont{Yurkevich and Lerner}(1999)}]{Yurkevich99}
\bibinfo{author}{\bibfnamefont{I.~V.} \bibnamefont{Yurkevich}}
  \bibnamefont{and} \bibinfo{author}{\bibfnamefont{I.~V.}
  \bibnamefont{Lerner}}, \bibinfo{journal}{Phys. Rev. B}
  \textbf{\bibinfo{volume}{60}}, \bibinfo{pages}{3955} (\bibinfo{year}{1999}).

\bibitem[{\citenamefont{Altland and Kamenev}(2000)}]{Altland00}
\bibinfo{author}{\bibfnamefont{A.}~\bibnamefont{Altland}} \bibnamefont{and}
  \bibinfo{author}{\bibfnamefont{A.}~\bibnamefont{Kamenev}},
  \bibinfo{journal}{Phys. Rev. Lett.} \textbf{\bibinfo{volume}{85}},
  \bibinfo{pages}{5615} (\bibinfo{year}{2000}).

\bibitem[{\citenamefont{Kanzieper}(2002)}]{Kanzieper02}
\bibinfo{author}{\bibfnamefont{E.}~\bibnamefont{Kanzieper}},
  \bibinfo{journal}{Phys. Rev. Lett.} \textbf{\bibinfo{volume}{89}},
  \bibinfo{pages}{250201} (\bibinfo{year}{2002}).

\bibitem[{\citenamefont{Altshuler and Aronov}()}]{Altshuler85}
\bibinfo{author}{\bibfnamefont{B.~L.} \bibnamefont{Altshuler}}
  \bibnamefont{and} \bibinfo{author}{\bibfnamefont{A.~G.}
  \bibnamefont{Aronov}}, \bibinfo{howpublished}{{\it Electron-Electron
  Interactions in Disordered Conductors} in {\it Electron-Electron Interactions
  in Disordered Systems}, edited by A. J. Efros and M. Pollak (Elsevier,
  Amsterdam, 1985)}.

\bibitem[{\citenamefont{Negele and Orland}(1988)}]{Negele88}
\bibinfo{author}{\bibfnamefont{J.~W.} \bibnamefont{Negele}} \bibnamefont{and}
  \bibinfo{author}{\bibfnamefont{H.}~\bibnamefont{Orland}},
  \emph{\bibinfo{title}{Quantum Many-Particle Systems}}
  (\bibinfo{publisher}{Addison-Wesley}, \bibinfo{address}{Redwood City,
  CA}, \bibinfo{year}{1988}).

\bibitem[{\citenamefont{Finkel'stein}()}]{Finkelstein90}
\bibinfo{author}{\bibfnamefont{A.~M.} \bibnamefont{Finkel'stein}},
  \bibinfo{howpublished}{in {\it Electron Liquid in Disordered Conductors},
  edited by I. M. Khalatnikov, Soviet Scientific Reviews Vol. 14 (Harwood,
  London, 1990)}.

\bibitem[{\citenamefont{Belitz and Kirkpatrick}(1994)}]{Belitz94}
\bibinfo{author}{\bibfnamefont{D.}~\bibnamefont{Belitz}} \bibnamefont{and}
  \bibinfo{author}{\bibfnamefont{T.}~\bibnamefont{Kirkpatrick}},
  \bibinfo{journal}{Rev. Mod. Phys.} \textbf{\bibinfo{volume}{66}},
  \bibinfo{pages}{261} (\bibinfo{year}{1994}).

\bibitem[{\citenamefont{Castellani et~al.}(1984)\citenamefont{Castellani, Di{
  }Castro, Lee, and Ma}}]{Castellani83}
\bibinfo{author}{\bibfnamefont{C.}~\bibnamefont{Castellani}},
  \bibinfo{author}{\bibfnamefont{C.}~\bibnamefont{Di{ }Castro}},
  \bibinfo{author}{\bibfnamefont{P.}~\bibnamefont{Lee}}, \bibnamefont{and}
  \bibinfo{author}{\bibfnamefont{M.}~\bibnamefont{Ma}}, \bibinfo{journal}{Phys.
  Rev. B} \textbf{\bibinfo{volume}{30}}, \bibinfo{pages}{527}
  (\bibinfo{year}{1984}).

\bibitem[{\citenamefont{Aleiner et~al.}(1999)\citenamefont{Aleiner, Altshuler,
  and Gershenson}}]{Aleiner99}
\bibinfo{author}{\bibfnamefont{I.~L.} \bibnamefont{Aleiner}},
  \bibinfo{author}{\bibfnamefont{B.~L.} \bibnamefont{Altshuler}},
  \bibnamefont{and} \bibinfo{author}{\bibfnamefont{M.~E.}
  \bibnamefont{Gershenson}}, \bibinfo{journal}{Waves Random Media}
  \textbf{\bibinfo{volume}{9}}, \bibinfo{pages}{201} (\bibinfo{year}{1999}).

\bibitem[{\citenamefont{Di{ }Castro and Raimondi}()}]{DiCastro03}
\bibinfo{author}{\bibfnamefont{C.}~\bibnamefont{Di{ }Castro}}
  \bibnamefont{and} \bibinfo{author}{\bibfnamefont{R.}~\bibnamefont{Raimondi}},
  \bibinfo{howpublished}{in {\it Quantum Phenomena in Mesoscopic physics},
  edited by B. Altshuler, V. Tognetti, and A. Tagliacozzo, Proceedings of the
  International School of Physics Enrico Fermi, Vol. 151 (IOS, Amsterdam,
  2003)}.

\bibitem[{\citenamefont{Itzykson and Zuber}(1980)}]{Itzykson80}
\bibinfo{author}{\bibfnamefont{C.}~\bibnamefont{Itzykson}} \bibnamefont{and}
  \bibinfo{author}{\bibfnamefont{J.~B.} \bibnamefont{Zuber}},
  \bibinfo{journal}{J. Math. Phys.} \textbf{\bibinfo{volume}{21}},
  \bibinfo{pages}{411} (\bibinfo{year}{1980}).

\bibitem[{\citenamefont{Harish-Chandra}(1957)}]{Harish57}
\bibinfo{author}{\bibnamefont{Harish-Chandra}}, \bibinfo{journal}{Am. J. Math}
  \textbf{\bibinfo{volume}{79}}, \bibinfo{pages}{87} (\bibinfo{year}{1957}).

\bibitem[{\citenamefont{Guhr and Wettig}(1996)}]{Guhr96}
\bibinfo{author}{\bibfnamefont{T.}~\bibnamefont{Guhr}} \bibnamefont{and}
  \bibinfo{author}{\bibfnamefont{T.}~\bibnamefont{Wettig}},
  \bibinfo{journal}{J. Math. Phys.} \textbf{\bibinfo{volume}{37}},
  \bibinfo{pages}{6395} (\bibinfo{year}{1996}).

\end{thebibliography}

\end{document}